\documentclass[runningheads,envcountsame,a4paper,final]{article}
\newcommand{\class}{0}

\usepackage[utf8]{inputenc}
\usepackage{lmodern}
\usepackage[T1]{fontenc}

\usepackage{amsmath}
\usepackage{amsthm,amssymb}
\usepackage{ebproof}
\usepackage{mathtools}
\usepackage{thmtools,thm-restate}
\usepackage{cancel}

\usepackage{subcaption}

\usepackage[table]{xcolor}

\usepackage[enable]{long}

\usepackage{hyperref}
\usepackage[capitalise,noabbrev]{cleveref} %
\crefalias{premise}{enumi}
\crefname{premise}{Premise}{Premises}
\crefalias{conclusion}{enumi}
\crefname{conclusion}{Conclusion}{Conclusions}
\crefalias{case}{enumi}
\crefname{case}{Case}{Cases}
\crefalias{step}{enumi}
\crefname{step}{Step}{Steps}

\usepackage[USenglish]{babel}
\usepackage{csquotes} %

\usepackage{enumitem}
\setlist[enumerate,1]{label=\arabic*.,ref=\arabic*,leftmargin=8pt}
\setlist[enumerate,2]{label*=\arabic*.,leftmargin=8pt}
\setlist[enumerate,3]{label*=\arabic*.,leftmargin=8pt}
\setlist[enumerate,4]{label*=\arabic*.,leftmargin=8pt}
\usepackage[oldenum,olditem]{paralist} %

\usepackage{tikz}
\usetikzlibrary{calc,shapes.misc,positioning,patterns,patterns.meta,fadings,arrows.meta}

\usepackage{microtype}

\usepackage{comment}
\usepackage{fancydraft}
\usepackage{notation}

\makeatletter
\AtBeginDocument{%
\@ifpackageloaded{hyperref}{%
\def\doi#1{\href{https://doi.org/\detokenize{#1}}{\texttt{doi:\detokenize{#1}}}}}{
\def\doi#1{https://doi.org/\detokenize{#1}}}
}
\makeatother

\title{Symbolic Model Construction for Saturated Constrained Horn Clauses}

\newcommand{\interp}{\mathcal{I}}
\newcommand{\jnterp}{\mathcal{J}}
\newcommand{\interpn}{\interp_{N}}
\newcommand{\sinterp}{\mathcal{S}}
\newcommand{\sjnterp}{\mathcal{R}}
\newcommand{\sinterpn}{\sinterp_{N}}
\newcommand{\sinterps}{\mathfrak{S}}
\newcommand{\ibot}{{\mathcal{I}_{\bot}}} %
\newcommand{\itop}{{\mathcal{I}_{\top}}} %
\newcommand{\isbot}{{\sinterp_{\bot}}} %
\newcommand{\istop}{{\sinterp_{\top}}} %

\newcommand{\LQA}{\ensuremath{\operatorname{LQA}}}
\newcommand{\laeq}{\approx}
\newcommand{\nlaeq}{\not\laeq}

\newcommand{\proofthmhornproducer}{%
\begin{proof}
Assume $\sinterp_{\preceq P} \models P(\vec{a})$.

Firstly, we argue that $\vec{a} \in P^{\Delta_P}$:
For the construction of $\sinterp_{\prec P}$ only predicates $Q$ s.t.\
$Q \prec P$ are considered, thus $P^{\sinterp_{\prec P}} = \emptyset$.
By definition of $\sinterp_{\preceq P}$ as
$\sinterp_{\preceq P}$ as the union of $\sinterp_{\prec P}$ and $\Delta_P$
and thus $P^{\sinterp_{\preceq P}} = P^{\sinterp_{\prec P}} \cup P^{\Delta_P}$.
Together with $P^{\sinterp_{\prec P}} = \emptyset$ we have
$P^{\sinterp_{\preceq P}} = P^{\Delta_P}$,
hence $\vec{a} \in P^{\Delta_P}$.

Secondly, we argue for the existence of $\ccx{C}$:
$\Delta_{P}$ is defined as the union of $\Delta_{P}^{\cc{\Lambda}{D' \lor P(\ast)}}$
where $\cc{\Lambda}{D' \lor P(\ast)} \in N$.
Thus
there must be
at least one $\delta(\sinterp_{\prec P}, \cc{\Lambda}{D' \lor L})$ with
$L = P(\vec{y})$, $D' \prec L$, $\sinterp_{\prec P} \nmodels \cc{\Lambda}{D' \lor L}$,
s.t.\ $\vec{a} \in P^{\delta(\sinterp_{\prec P}, \cc{\Lambda}{D' \lor L})}$.
We identify $\cc{\Lambda}{D' \lor L}$ with $\ccx{C}$.

Thirdly, we construct $\tau$,
and show
$\sinterp_{\prec P} \nmodels (\ccx{C})\tau$
by unfolding the interpretation of $P$ by $\sinterp_{\prec P}$:
Let $\beta_x = \{ \seqi{x_\i \mapsto a_\i} \}$.
From $\vec{a} \in \Delta_{P}^{\cc{\Lambda_C}{C}}$
we know that $\beta_x$ is a solution for
$(\pi (\{\vec{y}\}, \bigwedge_{\lambda \in \Lambda_C} \lambda \land \bigwedge_{i=1}^m (P_i^{\sinterp_{\prec P}})\sigma_i)) \sigma \wedge \curlyvee(\vec{y}, \vec{x})$.
Thus, $\sigma\beta_x$ is a solution for
$\pi (\{\vec{y}\}, \bigwedge_{\lambda \in \Lambda_C} \lambda \land \bigwedge_{i=1}^m (P_i^{\sinterp_{\prec P}})\sigma_i)$.
Let
$$\gamma \defsym \left\{ z \mapsto c_z \ \middle\vert \ z \in \vars(C) \setminus \vars(L) \right\}$$
be a witness for the existential quantifiers introduced by
$\pi$,
i.e.\
$\models (\bigwedge_{\lambda \in \Lambda_C} \lambda \land \bigwedge_{i=1}^m (P_i^\sinterp)\sigma_i)\gamma\sigma\beta_x$.
The composition $\tau \defsym \gamma\sigma\beta_x$
then maps all variables from
$C$ and is a solution for
$\bigwedge_{\lambda \in \Lambda_C} \lambda \land \bigwedge_{i=1}^m (P_i^\sinterp)\sigma_i$.
Consequently, $\tau$ is a solution for all conjuncts:
\begin{itemize}
\item{%
$\models \lambda\tau$ for all $\lambda \in \Lambda_C$ immediately
gives $\models \Lambda_C \tau$. 
}
\item{%
$\models (P_i^\sinterp \sigma_i)\tau$ for all $1 \leq i \leq m$
witnesses $\sinterp_{\prec P} \nmodels L_i\tau$, because
the polarity of $P_i^{\sinterp_{\prec P}}$ is opposite of
the polarity of $L_i$ by definition of $\delta$.
}
\end{itemize}
We have $P^{\sinterp_{\prec P}} = \emptyset$,
and that $L$ is positive, thus $\sinterp_{\prec P} \nmodels L$.
Together with the two above facts we arrive at $\sinterp_{\prec P} \nmodels (\ccx{C})\tau$.

To see that $P(\vec{y})\tau = P(\vec{a})$,
consider that
$(y_i)\sigma = x_i$, $(x_i)\beta_x = a_i$, and
$\gamma$ has no effect on $\vec{y}$ by definition.
Thus $(y_i)\sigma\beta_x = (y_i)\tau = a_i$.
In case there are two (or more, analoguously)
variables $y_i$, $y_j$ in $\vec{y}$
where $y_i = y_j$ and $i < j$,
$a_i = a_j$ is guaranteed:
$(y_i)\tau = (x)\beta_x = a_i$ directly by definition of $\sigma$ and $\beta_x$.
$y_j$ is not in the domain of $\sigma$, however
the equalities generated by $\curlyvee(\vec{y}, \vec{x})$
ensure that $(x_i)\beta_x = (x_j)\beta_x$.
\end{proof}
}

\newcommand{\proofthmhornproduces}{
\begin{proof}
Let $\ccx{C}$ where
$C = C' \lor P(\vec{y})$ and
$C' \prec P(\vec{y})$.
Let $\tau$ be a grounding substitution for $\ccx{C}$.
Assume $\sinterp_{\prec P} \nmodels (\ccx{C})\tau$.
This implies $\sinterp_{\prec P} \nmodels \ccx{C}$,
thus $\Delta_P^{\ccx{C}} = \delta(\sinterp_{\prec P}, \ccx{C})$.
Let
$\beta_\tau \defsym \{ x_i \mapsto y_i \tau \mid 1 \leq i \leq n \}$
and
$\phi = \bigwedge_{\lambda \in \Lambda_C} \lambda \land \bigwedge_{i=1}^m (P_i^{\sinterp_{\prec P}})\sigma_i$.
It remains to show that
$\Delta_P \models P(\vec{y})\tau$, i.e.\
$(\vec{y})\tau \in P^{\delta(\sinterp_{\prec P}, \ccx{C})}$, i.e.\
$\models ((\pi (\{\vec{y}\}, \phi)) \sigma \wedge \curlyvee(\vec{y}, \vec{x}))\beta_\tau$.
We proceed in two steps, one per conjunct:
\begin{enumerate}
\item{To see $\models (\curlyvee(\vec{y}, \vec{x}))\beta_\tau$:
Let $1 \leq i < j \leq n = \size{\vec{y}} = \size{\vec{x}}$.
By definition of $\curlyvee$,
for each conjunct $x_i \laeq x_j$ in $\curlyvee(\vec{y}, \vec{x})$,
there are two variables $y_i = y_j$ from $\vec{y}$.
The conjunct $x_i \laeq x_j$ is satisfied by
$\beta_\tau$, since $\beta_\tau$ and $\tau$ are functions,
therefore $(x_i)\beta_\tau = (x_j)\beta_\tau$.
}
\item{To see $\models (\pi (\{\vec{y}\}, \phi)) \sigma\beta_\tau$:
From $\sinterp_{\prec P} \nmodels (\ccx{C})\tau$,
we know $\models \Lambda_C \tau$,
thus $\models \lambda \tau$ for all $\lambda \in \Lambda_C$,
and
that $\sinterp_{\prec P} \nmodels (L_i) \tau$
for all literals $L_i$ from $C'$.
Given that $P_i^{\sinterp_{\prec P}}$ encodes $\sinterp_{\prec P} \nmodels L_i$
it follows that $\models P_i^{\sinterp_{\prec P}}\sigma_i\tau$ for all $1 \leq i \leq m$.

Thus, we may construct a potential witness for all
existential quantifiers introduced by the application of $\pi$
based on $\tau$:
$\gamma_\tau \defsym \{ z_i \mapsto z_i \tau \mid z_i \in \vars(\ccx{C}) \setminus \{ \vec{y} \} \}$.

By definition of $\sigma$ and $\beta_\tau$,
we have $(y_i)\sigma\beta_\tau = (y_i)\tau$
for all $y_i$ in $\vec{y}$.
Thus
$(\phi)\gamma_\tau \sigma \beta_\tau$
is equivalent to
$(\phi)\tau$.
Again, reasoning from $\sinterp_{\prec P} \nmodels (\ccx{C})\tau$
we arrive at $\models (\phi)\tau$,
thus
$\models (\pi (\{\vec{y}\}, \phi)) \sigma \beta_\tau$
with witness $\gamma_\tau$.
}
\end{enumerate}
Hence
$\models ((\pi (\{\vec{y}\}, \phi)) \sigma \wedge \curlyvee(\vec{y}, \vec{x}))\beta_\tau$,
thus
$\Delta_P^{\ccx{C}} \models \ccx{C}$,
and by definition of $\sinterp_{\preceq P}$,
also $\sinterp_{\preceq P} \models \ccx{C}$.
\end{proof}
}

\newcommand{\proofthmhorngeq}{%
\begin{proof}
Corollary of \cref{thm:horn:keep}.
\end{proof}
}

\newcommand{\proofthmhornmodelorproductivemodeltwo}{%
\begin{proof}
Follows from \cref{thm:horn:geq} and the fact that $\interpn = \sinterp_{\preceq Q}$
where $Q$ is the $\prec$-maximal predicate symbol in $N$.
\end{proof}
}

\newcommand{\proofthmhornmain}{%
\begin{proof}
By \cref{thm:horn:saturatedmodelorproductive} and \cref{thm:horn:modelorproductivemodel2}.
\end{proof}
}

\newcommand{\proofthmhornunsat}{%
\begin{proof}
Similar to \cref{thm:horn:saturatedmodelorproductive:proof:litneg} in the proof of \cref{thm:horn:saturatedmodelorproductive}.
\end{proof}
}

\newcommand{\proofthmhornlm}{%
\begin{proof}
Assume, towards a contradiction, that $\interpn$ is not the least model of $N$, i.e.\
there exists an interpretation $\interp$
such that $\interp \subset \interpn$ and  $\interp \models N$.
Since $\interp \subset \interpn$,
there is a predicate symbol $P$
and a point $\vec{a}$
such that $\vec{a} \in P^{\interpn}$, i.e.\ $\interpn \models P(\vec{a})$
but $\vec{a} \not \in P^{\interp}$, i.e.\ $\interp \nmodels P(\vec{a})$.
Assume, w.l.o.g., that $P$ is minimal, i.e.\
$Q^{\interpn} = Q^{\interp}$ for all $Q \prec P$.
By \cref{thm:horn:producer} from $\interpn \models P(\vec{a})$
it follows that there is a clause $\ccx{C} \in N$ such that
$C = C' \lor P(\vec{y})$,
$C' \prec P(\vec{y})$,
$\tau$ is a grounding for $\ccx{C}$,
$P(\vec{a}) = P(\vec{y})\tau$, and
$\sinterp_{\prec P} \nmodels (\ccx{C})\tau$.
From $\sinterp_{\prec P} \nmodels (\ccx{C})\tau$ we know that
$\sinterp_{\prec P} \nmodels (\cc{\Lambda_C}{C'})\tau$ and by
\cref{thm:horn:keep} we have $\interpn \nmodels (\cc{\Lambda_C}{C'})\tau$.
Since $C' \prec P(\vec{y})$ and by minimality of $P$,
we know that $\interpn$ and $\interp$
agree on $\cc{\Lambda_C}{C'}$, i.e.\
$\interp \nmodels (\cc{\Lambda}{C'})\tau$.
However, $\interp \models N$,
which implies $\interp \models (\ccx{C})\tau$,
requires $\interp \models P(a)$ which
contradicts the assumption $\interp \nmodels P(\vec{a})$.
\end{proof}
}

\newcommand{\proofthmhorncomplexity}{%
\begin{proof}
We proceed in two steps:
\begin{enumerate}
\item{\label[step]{thm:horn:complexity:proof:size}%
We argue that, for $P \in \preds(N)$, $\size{\atoms(P^{\sinterpn})}$ is in
$O(n^p \cdot m^{p-1} \cdot (l+a^2))$
and that $\size{\quants(P^{\sinterpn})}$ is in $O(n^{p-1} \cdot m^{p-1} \cdot q)$:
Induct over the predicates $\seq[p]{P}[1][\prec] \in \preds(N)$, along the order $\prec$, following \cref{def:modelconstruction}.
\begin{enumerate}
\item{\label[case]{thm:horn:complexity:proof:size:base}%
Base.
Since $P_1$ is $\prec$-minimal, we have $\sinterp_{\prec P_1} = \isbot$.
Thus, by definition of $\sinterp_{\preceq P_1}$ and union of symbolic interpretations, we have $P_1^{\sinterp_{\preceq P_1}} = \bot \lor \Delta_{P_1}$. 
Thus, $\atoms(P_1^{\sinterp_{\preceq P_1}}) = \atoms(\Delta_{P_1})$.
Next, unfolding the definition of $\Delta_{P_1}$, observe that it is defined as a union of symbolic interpretations, and the operands correspond to some clause $\cc{\Lambda}{C}$ where $C = P_1(\ast) \lor C'$, i.e., clauses where $P_1$ occurs positively.
Consider the definition of $\Delta_{P_1}^{\cc{\Lambda}{C}}$.
If $P_1(\ast)$ is not maximal in $C$, then $P_1^{\Delta_{P_1}^{\cc{\Lambda}{C}}} = \bot$, which contains no atoms.
Otherwise, i.e., $P_1(\ast)$ is maximal, then $P_1^{\Delta_{P_1}^{\cc{\Lambda}{C}}} = \delta(\sinterp_{\prec P_1}, \cc{\Lambda}{C})$.
Since the number of clauses where $P_1(\ast)$ is maximal is in $O(n)$, $\size{\atoms(\Delta_{P_1})}$ is in $O(n \cdot \size{\atoms(\delta(\sinterp_{\prec P_1}, \cc{\Lambda}{C}))})$, because $\Delta_{P_1}$ is just a disjunction over these formulas (and possibly multiple other disjuncts that are $\bot$) according to the definition of union for symbolic interpretatons.
Next, we argue the number of atoms and quantifiers for each of these $n$ disjuncts of the form $\delta(\sinterp_{\prec P_1}, \cc{\Lambda}{C})$ separately:
\begin{enumerate}
\item{$\size{\atoms(\delta(\sinterp_{\prec P_1}, \cc{\Lambda}{C}))}$ is bounded by $l + a^2$:
Consider the definition of $\delta$. For the interpretation of $P_1$, it results in a conjunction.
The first $l$ elements of the conjunction directly correspond to the elements of $\Lambda$,
and since the number of atoms in $\Lambda$ is bounded by $l$, this gives the first summand.
Note that the second block of conjuncts, arising from conjunctively combining $P^{\sinterp_{\prec P_1}}_i$ are all empty,
since $\sinterp_{\prec P_1} = \isbot$.
To see that the number of conjuncts generated by $\curlyvee$ is bounded by $a^2$,
assume that $P_1$ is of the maximal arity $a$, and all arguments of the literal $P_1(\ast)$ are the same.
Then, $\curlyvee$ will generate $a^2$ equality atoms among these $a$ variables.
}
\item{$\size{\quants(\delta(\sinterp_{\prec P_1}, \cc{\Lambda}{C}))}$ is bound by $q$:
Consider the definition of $\pi$.
Given a formula over the variables $\vars(\cc{\Lambda}{C})$,
it generates $\size{\vars(\cc{\Lambda}{C}) \setminus \vars(P_1(\ast))}$ quantifiers.
This number is at most $q$.
}
\end{enumerate}
Hence, $\size{\atoms(P_1^{\sinterp_{\preceq P_1}})}$ is in $O(n \cdot (l + a^2))$ and $\size{\quants(P_1^{\sinterp_{\preceq P_1}})}$ is in $O(n \cdot q)$.
}
\item{Step.
For the same reason as in \cref{thm:horn:complexity:proof:size:base}, $P_{i+1}^{\sinterp_{\preceq P_{i+1}}}$ is a disjunction over at most $n$ subformulas, one for each clause with maximal predicate $P_{i+1}$.
We argue the number of atoms and quantifiers for each of these $n$ disjuncts of the form $\delta(\sinterp_{\prec P_{i+1}}, \cc{\Lambda}{C})$ separately:
\begin{enumerate}
\item{$\size{\atoms(\delta(\sinterp_{\prec P_{i+1}}, \cc{\Lambda}{C}))}$ is bound by $l + a^2 + m \cdot \max_{k \leq i}\size{\atoms(P_k^{\sinterp_{\prec P_{i+1}}})}$:
The summands $l$ and $a^2$ are justified like in the base.
The additional summand is motivated by fact that, in contrast to the base, now $\sinterp_{\prec P_{i+1}} = \sinterp_{\preceq P_i} \neq \isbot$.
By the (strong) induction hypothesis, $\size{\atoms(P_i^{\sinterp_{\preceq P_i}})}$ is in $O(n^i \cdot m^{i-1} \cdot (l + a^2))$ for all $1 \leq i$.
We substitute using the hypothesis to get $\size{\atoms(P_{i+1}^{\sinterp_{\preceq P_{i+1}}})}$ in $O(n \cdot (l + a^2 + m \cdot (n^i \cdot m^{i-1} \cdot (l + a^2))))$.
We expand to $O(\underline{n^{i + 1} m^i} \cdot a^2 + \underline{n^{i+1} m^i} \cdot l + \cancel{n a^2} + \cancel{n l})$, drop the non-exponential summands and factor, which yields $\size{\atoms(P_{i+1}^{\sinterp_{\preceq P_{i+1}}})}$ in $O(n^{i+1} m^i (a^2 + l))$.
}
\item{$\size{\quants(\delta(\sinterp_{\prec P_{i+1}}, \cc{\Lambda}{C}))}$ is bound by $q + m \cdot \max_{k \leq i}\size{\quants(P_k^{\sinterp_{P_i}})}$:
The summand $q$ is justified like in the base case.
Like in the previous case, the additional summand $m \cdot \max_{k \leq i}\size{\quants(P_k^{\sinterp_{P_i}})}$ motivated by the fact that, in contrast to the base case, now $\sinterp_{\prec P_{i+1}} = \sinterp_{\preceq P_i} \neq \isbot$.
For each of the literals in $C$ that refer to a predicate symbol $P_{j}$ with $j \leq i$, as many quantifiers as there are in $P_{j}^{\sinterp_{\prec P_{i+1}}}$ are generated by $\delta$. 
By the (strong) induction hypothesis, $\size{\quants(P_i^{\sinterp_{\preceq P_i}})}$ is in $O(n^i \cdot m^{i-1} \cdot q)$ for all $1 \leq i$.
We substitute using the hypothesis to get $\size{\quants(P_{i+1}^{\sinterp_{\preceq P_{i+1}}})}$ in $O(n \cdot (q \cdot m \cdot (n^i \cdot m^{i-1} \cdot q)))$.
After expansion and simplification, this yields that $\size{\quants{P_{i+1}^{\preceq P_{i+1}}}}$ is in $O(n^{i+1} m^i q)$.
}
\end{enumerate}
Note that $\size{\atoms(P_i^{\sinterp_{\preceq P_i}})} = \size{\atoms(P_i^{\sinterpn})}$ (and the same is true for the number of quantifiers), which can be seen by \cref{thm:horn:keep}.
}
\end{enumerate}
}
\item{%
We argue that by applying quantifier elimination to all $P^{\sinterpn}$ for $P \in \preds(N)$, we obtain an $\sinterpn'$ from $\sinterpn$, which represents an equivalent interpretation, and we arrive at the conclusion. Let $e(q, n) \defsym n^q$ denote the number of atoms after eliminating $q$ quantifiers from the top of a formula with $n$ atoms (see \cite{DBLP:journals/cj/LoosW93,Cooper:72a}).
The inductive argument is very similar to \cref{thm:horn:complexity:proof:size}, just note that we eliminate quantifiers at the level of the disjuncts that form $\Delta$.
\begin{enumerate}
\item{Base.
For $P_1$ we have max.\ $q$ quantifiers for each of the (at most) $n$ clauses with $P_1$ as maximal predicate. Thus, $O(\size{\atoms(P_1^{\sinterp_{\preceq P_1}})}) = O(n \cdot e(q, l + a^2)) = O(n \cdot (l + a^2)^q)$.
}
\item{Step.
By the induction hypothesis:
\begin{align*}
O\big(\size{\atoms(P_i^{\sinterp_{\preceq P_i}})}\big)
=\ & O(n^{q^{i-1} + q^{i-2} + \dots + 1} \cdot m^{q^{i-1} + q^{i-2} + \dots + q} \cdot (l + a^2)^{q^i}) \\
=\ & O(n^{2 \cdot q^{i-1}} \cdot m^{2 \cdot q^{i-1}} \cdot (l + a^2)^{q^i})
\end{align*}

For $P_{i+1}$ we still have max. $q$ quantifiers for each of the $n$ clauses with $P_{i+1}$ as maximal predicate.
\begin{align*}
   & O\big(\size{\atoms(P_{i+1}^{\sinterp_{\preceq P_{i+1}}})}\big) \\
=\ & n \cdot e(q,(l + a^2 + m \cdot O((\size{\atoms(P_{i}^{\sinterp_{\preceq P_i}})})))) \\
=\ & O\big(n \cdot (l + a^2 + m \cdot n^{q^{i-1} + q^{i-2} + \dots + 1} \cdot m^{q^{i-1} + q^{i-2} + \dots + q} \cdot (l + a^2)^{q^i})^q\big) \\
\intertext{(We drop non-dominating summands $l + a^2$.)}
=\ & O\big(n \cdot m^q \cdot n^{q^i + q^{i-1} + \dots + q} \cdot m^{q^{i} + q^{i-1} + \dots + q^2} \cdot (l + a^2)^{q^{i+1}}\big) \\
=\ & O\big(n^{q^i + q^{i-1} + \dots + q + 1} \cdot m^{q^{i} + q^{i-1} + \dots + q^2 + q} \cdot (l + a^2)^{q^{i+1}}\big) \\
=\ & O(n^{2 \cdot q^{i}} \cdot m^{2 \cdot q^{i}} \cdot (l+a^2)^{q^{i+1}})
\end{align*}
}
\end{enumerate}
}
\end{enumerate}
\end{proof}
}

\newcommand{\proofthmhorneffective}{%
\begin{proof}
Corollary of \cref{thm:horn:complexity}. The construction terminates, since it computes $\sinterpn$ from $N$ (which are both finite) by a simple recursion (cf.\ \cref{def:modelconstruction}).
\end{proof}
} % end input ./proof.tex
\usepackage{myappendix}
\usepackage{authblk}

\newtheorem{theorem}{Theorem}{\bfseries}{\itshape}
\newtheorem{example}[theorem]{Example}{\itshape}{\rmfamily}

\newtheorem{definition}[theorem]{Definition}{\bfseries}{\itshape}
\newtheorem{proposition}[theorem]{Proposition}{\bfseries}{\itshape}
{\bfseries}{\itshape}
{\bfseries}{\itshape}

\date{2023-07-18}

\author{Martin Bromberger}
\author[1,2]{Lorenz Leutgeb}
\author{Christoph Weidenbach}

\affil{Max~Planck Institute~for Informatics, Saarland~Informatics~Campus, Saarbr\"ucken,~Germany\\\texttt{\{mbromber,lorenz,weidenb\}@mpi-inf.mpg.de}}
\affil[2]{Graduate~School~of~Computer~Science, Saarland~Informatics~Campus, Saarbr\"ucken, Germany}
\begin{document}

\lng{}[
\setlength{\belowdisplayskip}{0.5\belowdisplayskip}
\setlength{\abovedisplayskip}{0.5\abovedisplayskip}
]

\maketitle
\begin{abstract}%
%
% start input ./abstract.tex
%

Clause sets saturated by hierarchic ordered resolution do not
offer a model representation that can be effectively queried, in general. 
They only offer the guarantee
of the existence of a model.
We present an effective symbolic model construction for 
saturated constrained Horn clauses.
Constraints are in linear arithmetic, the first-order
part is restricted to a function-free language. The model
is constructed in 
finite time, and non-ground clauses can be effectively evaluated with respect to the model.
Furthermore, we prove that our model construction produces the least model.
 % end input ./abstract.tex
 \end{abstract}

%
% start input ./intro.tex
\section{Introduction}\label{sec:intro}%

Constrained Horn Clauses (CHCs) combine logical formulas with constraints
over various domains, e.g.\
linear real arithmetic,
linear integer arithmetic,
equalities of uninterpreted functions
\cite{DBLP:journals/tplp/AngelisFGHPP22}.
This formalism has gained widespread attention in recent years due to its applications in a
variety of fields, including
program analysis and verification: safety, liveness, and termination~\cite{DBLP:journals/toplas/SpotoMP10,DBLP:conf/cav/FedyukovichZG18},
complexity and resource analysis~\cite{DBLP:journals/tplp/Lopez-GarciaDKL18},
intermediate representation~\cite{DBLP:journals/tplp/GangeNSSS15},
and software testing~\cite{DBLP:journals/tplp/MesnardPV20}.
Technical controls, so called \emph{Supervisors}, like an electronic
engine control unit, or a lane change assistant in a car~\cite{DBLP:conf/frocos/BrombergerDFFKW21,DBLP:conf/tacas/BrombergerDFFGK22} can be
modelled, run, and proven safe.
Moreover, there exist many different approaches for reasoning in CHCs and associated first-order logic fragments extended with theories~\cite{DBLP:journals/tplp/AngelisFGHPP22,DBLP:conf/birthday/BjornerGMR15,DBLP:conf/pldi/GrebenshchikovLPR12,DBLP:conf/sat/HoderB12,DBLP:conf/cav/McMillan14,DBLP:conf/cav/KomuravelliGC14,DBLP:conf/kgc/BachmairGW93,DBLP:conf/lics/GanzingerN99,DBLP:conf/csl/KorovinV07,DBLP:conf/lpar/Rummer08,DBLP:conf/lpar/BaumgartnerFT08,DBLP:conf/vmcai/BrombergerFW21}.
Thus, CHCs are a powerful tool for reasoning about complex systems that
involve logical constraints, and they have been used to solve a wide range of
problems.

A failed proof attempt of some conjecture or undesired run points to a bug. In this case
investigation of the cause of the unexpected result or behavior is crucial.
Building a model of the situation that can then be effectively
queried is an important means towards a repair. 
However, some algorithms for CHCs, e.g.\ hierarchic superposition, 
which boils down to hierarchic ordered resolution in the context of CHCs,
do not return a model that can be effectively queried if a proof attempt fails, in general.
If so, queries are still restricted to ground clauses~\cite{DBLP:journals/jacm/BasinG01}.

The contribution of our paper can be seen as an extension for these 
saturation based algorithms that produces models and not just saturated clause sets.
In fact, we show how to build symbolic models out of any saturated CHC clause
set over linear arithmetic.
This fragment is equivalent to Horn clause sets of linear arithmetic combined with
the Bernays-Sch{\"o}nfinkel fragment.
Recall that although satisfiability in
this fragment is undecidable~\cite{Downey1972,DBLP:journals/corr/HorbachVW17}, in general, 
for a finitely saturated set we can construct such a representation in finite time.

Our models fulfill all important properties postulated 
in the literature for automated model building in first-order logic~\cite{DBLP:journals/igpl/FermullerL98,modelbuilding}.
First, they can be \emph{effectively constructed}, i.e., 
each model is represented by one linear arithmetic formula of finite size for each of its predicates and it can be constructed in finite time.
Second, they are \emph{unique}, i.e., 
the model representation specifies exactly one interpretation; in our case the least model.
Third, they can be \emph{effectively queried}, i.e., 
we provide decision procedures that evaluate whether an atom, clause, or formula is entailed/satisfied by the model.
Fourth, it is possible to \emph{test} the \emph{equivalence} of two models.
The approach we present does not exploit features of linear arithmetic beyond equality,
the existence of a well-founded order for the theories' universe, and
decidability of the theory.
The results may therefore be adapted to other constraint domains.
Model representation that can be effectively constructed and queried like ours are also called 
\emph{effective model representations}.
Moreover, our method is the first effective model construction approach
for ordered resolution (or its extension to superposition) that is based on saturation, 
goes beyond ground clauses, and includes theory constraints.
In the future, we plan to use this approach as the basis for a more general model construction approach that also works on
more expressive fragments of first-order logic modulo theories.

Our model construction is inspired by the model construction operator 
used in the proof for refutational completeness of hierarchic superposition~\cite{DBLP:phd/dnb/Kruglov13,DBLP:conf/birthday/0001W19,DBLP:journals/aaecc/BachmairGW94}.
The main difference is that the model construction operator from the refutational completeness proof is restricted 
to ground clauses and executed on the potentially infinite ground instances of the saturated clause set
(in addition to an infinite axiomatization of the background theory as ground clauses).
As a result, the model construction operator from the refutational completeness proof cannot 
effectively construct a model because iterating over a potentially infinite set means it may diverge.
Moreover, in contrast to our model construction, 
the original model operator cannot effectively evaluate non-ground atoms, clauses, or formulas.
It is, however, sufficient, to show the existence of a model
if the clause set is saturated and does not contain the empty clause~\cite{DBLP:phd/dnb/Kruglov13,DBLP:conf/birthday/0001W19,DBLP:journals/aaecc/BachmairGW94}.
In our version of the model construction operator, we managed to lift the restriction to ground clause sets 
by restricting the input logic to the Horn Bernays-Sch{\"o}nfinkel fragment instead of full first-order logic.
This enables us to define a strict propagation/production order for our non-ground clauses 
instead of just for ground clauses.
As a result, we can construct the model one clause at a time.

The paper is organized as follows.
In \cref{sec:prelim} we clarify notation and preliminaries.
The main contribution is presented in \cref{sec:horn}.
At the end of this section, we also explain how our models satisfy the postulates (see \cite[Section~5.1,~p.~234]{modelbuilding}) by Fermüller and Leitsch for automated model building.
We conclude in \cref{sec:conclusion}.
\ifnum\class=1
{Proofs were elided in favor of explanations and examples. An extended version, which includes proofs, can be found at \cite{DBLP:journals/corr/abs-2305-05064}.}
\fi
 % end input ./intro.tex
 %
% start input ./prelim.tex
\section{Preliminaries and Notation}\label{sec:prelim}

We briefly recall the basic logical formalisms and notations we build
upon~\cite{DBLP:conf/frocos/BrombergerDFFKW21}.
Our starting point is a standard first-order language with
\emph{variables} (denoted $x,y,z$),
\emph{predicates} (denoted $P, Q$) of some fixed \emph{arity},
and \emph{terms} (denoted $t, s$).
An \emph{atom} (denoted $A$) is an expression $P(\seq{t})$ for a predicate $P$ of arity $n = \arity(P)$.
When the terms $\seq{t}$ in $P(\seq{t})$ are not relevant in some context,
we also write $P(\ast)$.
A \emph{positive literal} is an atom $A$ and a \emph{negative literal} is a negated atom $\lnot A$.
We define $\comp(A)=\lnot A$, $\comp(\lnot A)=A$, $|A|=A$ and $|\lnot A|=A$.
Literals are usually denoted $L, K$.
We sometimes write literals as $[\lnot]P(\ast)$,
meaning that the sign of the literal is arbitrary,
often followed by a case distinction.
Formulas are defined in the usual way using quantifiers $\forall$, $\exists$ and
the boolean connectives (in order of decreasing binding strength)
$\neg$, $\lor$, $\land$, $\rightarrow$, and $\leftrightarrow$.
The logic we consider does not feature a first-order equality predicate.

A \emph{clause} (denoted $C, D$) is a universally closed disjunction of literals
$\seq{A}[1][\lor] \lor \seq[m]{\lnot B}[1][\lor]$.
We may equivalently write $\seq[m]{B}[1][\land] \to \seq{A}[1][\lor]$.
A clause is \emph{Horn} if it contains at most one positive literal, i.e.\ $n \leq 1$.
In \cref{sec:horn}, all clauses considered are Horn clauses.
If $Y$ is a term, formula, or a set thereof, $\vars(Y)$ denotes the set of all variables in $Y$,
and $Y$ is \emph{ground} if $\vars(Y) = \emptyset$.
Analogously, $\preds(Y)$ is the set of predicate symbols occurring in $Y$.

The \emph{\BSfull\ Clause Fragment} (\BS{}) in first-order logic consists of first-order clauses
where all terms are either variables or constants.
The \emph{Horn \BSfull\ Clause Fragment} (\HBS{}) is further restricted to Horn clauses.

A \emph{substitution} $\sigma$ is a function from variables to terms with a finite
domain \lng{$\dom(\sigma) = \{ x \mid x\sigma \neq x\}$} and
codomain \lng{$\cdom(\sigma) = \{x\sigma\mid x\in\dom(\sigma)\}$}.
We denote substitutions by $\sigma, \tau$.
The application of substitutions is often written postfix, as in $x\sigma$, and is homomorphically extended to
terms, atoms, literals, clauses, and quantifier-free formulas.
A substitution \lng{$\sigma$} is \emph{ground} if \lng{$\cdom(\sigma)$}[its codomain] is ground.
Let $Y$ denote some term, literal, clause, or clause set.
A substitution $\sigma$ is a \emph{grounding} for $Y$ if $Y\sigma$ is ground, and $Y\sigma$ is a
\emph{ground instance} of $Y$ in this case.
We denote by $\mGnd(Y)$ the set of all ground instances of $Y$.
The \emph{most general unifier} $\mMGU(Z_1,Z_2)$ of two terms/atoms/literals $Z_1$ and $Z_2$
is defined as usual, and we assume that it does not introduce fresh variables and is idempotent.

\subsection{Horn \BSfull\ with Linear Arithmetic}\label{sec:prelim:la}

The class $\HBS(\LRA)$ is the extension of the Horn \BSfull{} fragment with linear real arithmetic ($\LRA$).
Analogously, the classes $\HBS(\LQA)$ and $\HBS(\LIA)$ are the extensions of the Horn \BSfull{} fragment with 
linear rational arithmetic ($\LQA$) and linear integer arithmetic ($\LIA$), respectively.
The only difference between the three classes are the sort $\LA$ their variables and terms range over and 
the universe $\mathcal{U}$ over which their interpretations range.
As the names already imply $\LA = \LRA$ and $\mathcal{U} = \mathbb{R}$ for $\HBS(\LRA)$, $\LA = \LQA$ and $\mathcal{U} = \mathbb{Q}$ for 
$\HBS(\LQA)$, and $\LA = \LIA$ and $\mathcal{U} = \mathbb{Z}$ for $\HBS(\LIA)$. 
The results presented in this paper hold for all three classes and 
by $\HBS(\LA)$ we denote that we are talking about an arbitrary one of them.

Linear arithmetic terms are constructed from a set $\varset$ of \emph{variables}, %
the set of constants $c\in\mathbb{Q}$ (if in $\HBS(\LRA)$ or $\HBS(\LQA)$) or $c\in\mathbb{Z}$ (if in $\HBS(\LIA)$), and binary function symbols $+$ and $-$ (written infix).
Additionally, we allow multiplication $\cdot$ if one of the factors is a constant.
Multiplication only serves us as syntactic sugar to abbreviate other arithmetic terms, 
e.g., $x + x + x$ is abbreviated to $3 \cdot x$.
Atoms in $\HBS(\LA)$ are either \emph{first-order atoms} (e.g., $P(13,x)$) or
\emph{(linear) arithmetic atoms} (e.g., $x < 42$).
Arithmetic atoms are denoted by $\lambda$ and may use the predicates $\leq, <, \laeq, \nlaeq, >, \geq$,
which are written infix and have the expected fixed interpretation. 
We use $\laeq$ instead of $=$ to avoid confusion between equality in $\LA$ and equality on the meta level.
While we do not permit quantifiers in the syntax of clauses, the notion of symbolic interpretations that we will develop does require this, denoted as usual.
By $\atoms(Y)$/$\quants(Y)$ we denote the linear arithmetic atoms/quantifiers in a formula or set of formulas $Y$.
\emph{First-order literals} and related notation is defined as before.
\emph{Arithmetic literals} coincide with arithmetic atoms, since
the arithmetic predicates are closed under negation, e.g.,
$\lnot(x \geq 42)$ is equivalent to $x < 42$.

$\HBS(\LA)$ clauses are defined as for \HBS{} but using $\HBS(\LA)$ atoms.
We often write clauses in the form $\cc{\Lambda}{C}$ where $C$ is a clause solely built of free first-order literals
and $\Lambda$ is a multiset of $\LA$ atoms called the \emph{constraint} of the clause. 
A clause of the form $\cc{\Lambda}{C}$ is therefore also called a \emph{constrained clause}.
Since the interpretation of linear arithmetic relations is fixed,
we set $\preds(\cc{\Lambda}{C}) \defsym \preds(C)$.

The fragment we consider in \cref{sec:horn} is restricted
even further to \emph{abstracted} clauses:
For any clause $\cc{\Lambda}{C}$,
all terms in $C$ must be variables.
Put differently, we disallow any arithmetic function symbols,
including numerical constants, in $C$.
Variable abstraction, e.g.\ rewriting $\cc{x \geq 3}{P(x,1)}$
to $\cc{x \geq 3, y \laeq 1}{P(x,y)}$, is always possible.
Hence, the restriction to abstracted clauses is not a theoretical limitation,
but allows us to formulate our
model construction operator in a more concise way.
We assume abstracted clauses for theory development, but we prefer non-abstracted
clauses in examples for readability, e.g., a unit clause $P(3,5)$ is considered in the development
of the theory as the clause $\cc{x\laeq3, y\laeq5}{P(x,y)}$.

In contrast to other works, e.g.\ \cite{DBLP:conf/cade/BrombergerLW22},
we do not permit first-order constants, and consequently also no variables that range over the induced Herbrand universe.
All variables are arithmetic in the sense that they are interpreted by $\mathcal{U}$.
Since we only allow equalities in the arithmetic constraint, it is possible to simulate variables over first-order constants, by e.g.\ numbering them,
i.e.\ defining a bijection between $\mathbb{N}$ and constant symbols. So this again not a theoretical limitation.

\begin{comment}
A clause or clause set is \emph{abstracted} if its first-order literals contain only variables or first-order constants.
Every clause $C$ is equivalent to an abstracted clause that is obtained by replacing each non-variable arithmetic term $t$
that occurs in a first-order atom by a fresh variable $x$ while adding an arithmetic atom $x\nlaeq t$ to $C$.
\end{comment}

The semantics of $\cc{\Lambda}{C}$ is as follows:
$$\cc{\Lambda}{C} \quad \text{iff} \quad
\big(\bigwedge_{\lambda \in \Lambda}\lambda\big) \to C \quad \text{iff} \quad
\big(\bigvee_{\lambda \in \Lambda} \lnot \lambda\big) \vee C$$
For example, the clause $x >1 \lor y \nlaeq 5 \lor \lnot Q(x) \lor R(x, y)$
is also written $\cc{x\leq 1, y \laeq 5}{\lnot Q(x) \lor R(x, y)}$.
The negation $\lnot(\cc{\Lambda}{C})$ of a constrained clause $\cc{\Lambda}{C}$ where
$C = \seq{A}[1][\lor] \lor \seq[m]{\lnot B}[1][\lor]$
is thus equivalent to $(\bigwedge_{\lambda \in \Lambda} \lambda) \land \seq{\lnot A}[1][\land] \land \seq[m]{B}[1][\land]$.
Note that since the neutral element of conjunction is $\top$,
an empty constraint is thus valid, i.e.\ equivalent to true.
In analogy to the empty clause in settings without constraints,
we write $\square$ to mean any and all clauses $\cc{\Lambda}{\bot}$
where $\Lambda$ is satisfiable, which are all unsatisfiable.

An \emph{assignment} for a constraint $\Lambda$ is a substitution (denoted 
$\beta$) that maps all variables in $\vars(\Lambda)$ to values in $\mathcal{U}$. 
An assignment is a \emph{solution} for a constraint $\Lambda$ if all atoms $\lambda \in (\Lambda \beta)$ evaluate to true.
A constraint $\Lambda$ is \emph{satisfiable} if there exists a solution for $\Lambda$. 
Otherwise it is \emph{unsatisfiable}.
We assume \emph{pure} input clause sets because otherwise satisfiability is undecidable for impure $\HBS(\LA)$~\cite{DBLP:journals/corr/abs-2003-04627}.
This means the only constants of our sort $\LA$ are concrete rational numbers.
Irrational numbers are not allowed by the standard definition of the theory.
Fractions are not allowed if $\LA = \LIA$.
Satisfiability of pure $\HBS(\LA)$ clause sets is semi-decidable, e.g., using \emph{hierarchic superposition}~\cite{DBLP:journals/aaecc/BachmairGW94}
or \emph{SCL(T)}~\cite{DBLP:conf/vmcai/BrombergerFW21}.
Note that pure $\HBS(\LA)$ clauses correspond to \emph{constrained Horn clauses (CHCs)} with $\LA$ as background theory.

All arithmetic predicates and functions are interpreted in the usual way denoted by the interpretation $\sigval^{\LA}$.
An interpretation of $\HBS(\LA)$ coincides with $\sigval^{\LA}$ on arithmetic predicates and functions, and freely interprets
non-arithmetic predicates. For pure clause sets this is well-defined~\cite{DBLP:journals/aaecc/BachmairGW94}.
Logical satisfaction and entailment is defined as usual, and uses similar notation as for \HBS.

\begin{example}\label{ex:ta}
The clause
$\cc{y \geq 5,\,x' \laeq x + 1}{S_0(x,y) \to S_1(x', 0)}$
is part of a timed automaton with two clocks $x$ and $y$ modeled in $\HBS(\LA)$.
It represents a transition from state $S_0$ to state $S_1$ that can be traversed only if 
clock $y$ is at least $5$ and that resets $y$ to $0$ and increases $x$ by $1$.
\end{example}

\begin{comment}
Most algorithms for \BSfull{}, first-order logic, and beyond utilize resolution. 
The $\SCLT$ calculus for $\HBS(\LA)$ uses hierarchic resolution 
in order to learn from the conflicts it encounters during its search.
The hierarchic superposition calculus on the other hand derives new clauses via hierarchic resolution based on an ordering.
The goal is to either derive the empty clause or a saturation of the clause set, i.e., 
a state from which no new clauses can be derived.
Each of those algorithms must derive new clauses in order to progress, 
but their subroutines also get progressively slower as more clauses are derived.
In order to increase efficiency, it is necessary to eliminate clauses that are obsolete.
One measure that determines whether a clause is useful or not is \emph{redundancy}.
\end{comment}

\subsection{Ordering Literals and Clauses}

In order to define redundancy for constrained clauses, we need an \emph{order}: Let
$\prec_{\preds}$ be a total, well-founded, strict ordering on predicate symbols
and let $\prec_{\mathcal{U}}$ be a total, well-founded, strict ordering on the universe $\mathcal{U}$.
(Note that $\prec$ cannot be the standard ordering $<$ because it is not well-founded for $\mathbb{Z}$, $\mathbb{Q}$, or $\mathbb{R}$.
In the case of $\mathbb{R}$, the existence of such an order is even dependent on whether we assume the axiom of choice~\cite{Feferman1964}.)
We extend these orders step by step.
First, to atoms, i.e.,
$P(\vec{a}) \prec Q(\vec{b})$ if $P \prec_{\preds} Q$ or $P = Q$, $\vec{a}, \vec{b} \in \mathcal{U}^{\size{\vec{a}}}$, and $\vec{a} \prec_{\text{lex}} \vec{b}$,
where $\prec_{\text{lex}}$ is the lexicographic extension of $\prec_{\mathcal{U}}$.
Next, we extend the order to literals with a strict precedence on the predicate and the polarity, i.e.,
\begin{equation*}
P(\vec{t}) \prec \lnot P(\vec{s}) \prec Q(\vec{u}) \qquad \text{if $P \prec Q$}
\end{equation*}
independent of the arguments of the literals.
Then, take the multiset extension to order clauses.
To handle constrained clauses extend the relation
such that constraint literals (in our case arithmetic literals)
are always smaller than first-order literals.
We conflate the notation of all extensions into the symbol
$\prec$ and define $\preceq$ as the reflexive closure of $\prec$.
Note that $\prec$ is only total for ground atoms/literals/clauses,
which is sufficient for a hierarchic superposition order~\cite{DBLP:conf/birthday/0001W19}.

\begin{definition}[$\prec$-maximal Literal]\label{def:maxlit}
A literal $L$ is called \emph{$\prec$-maximal} in a clause $C$
if there exists a grounding substitution $\sigma$
for $C$,
such that
there is no different $L' \in C$ for which $L\sigma \prec L'\sigma$.
The literal $L$ is called \emph{strictly $\prec$-maximal}
if there is no different $L' \in C$ for which $L\sigma \preceq L'\sigma$.
\end{definition}

\begin{proposition}
If $\prec$ is a predicate-based ordering, $C$ is a Horn clause,
$C$ has a positive literal $L$, and $L$ is $\prec$-maximal in $C$, then $L$ is strictly $\prec$-maximal in $C$.
\end{proposition}

\begin{definition}[$\prec$-maximal Predicate in Clause]\label{def:maxpred}
A predicate symbol $P$ is called \emph{(strictly) $\prec$-maximal} in a clause $C$
if there is a literal $[\lnot]P(\ast) \in C$ that is (strictly) $\prec$-maximal in $C$.
\end{definition}

\begin{definition}
Let $N$ be a set of clauses, $\prec$ a clause ordering, $C$ a clause, and $P$ a predicate symbol.
Then $N^{\prec C} \defsym \{ C' \in N \mid C' \prec C \}$ and $N^{\preceq P} \defsym \{ C \in N \mid Q \text{ is $\prec$-maximal in } C \text{ and } Q \preceq P\}$.
\end{definition}

%
\begin{comment}
\begin{definition}[Signed Predicate Symbol]
A \emph{signed predicate symbol} is a predicate symbol,
optionally prefixed with negation.
Let $P$ be a predicate symbol,
then $P$ itself is the positive signed predicate symbol corresponding to $P$,
and $\lnot P$ is the negative signed predicate symbol corresponding to $P$.
\end{definition}

\begin{definition}[Signed Predicate Symbol of a Literal]
Let $L$ be a literal. If $L$ is of the form $P(\ast)$ then
its corresponding signed predicate symbol is $P$,
otherwise, i.e.\ if $L$ is of the form $\lnot P(\ast)$, then
its corresponding signed predicate symbol is $\lnot P$.
We denote the function that projects a literal to its corresponding
signed predicate symbol by $s$.
\end{definition}

\begin{definition}[$\prec$-maximal Signed Predicate Symbol]
Let $C = C' \lor L$ be a clause, and $C' \preceq L$,
then $s(L)$ is a $\prec$-maximal signed predicate symbol of $C$.
\end{definition}
\end{comment}

\subsection{Hierarchic Superposition, Redundancy and Saturation}

\begin{comment}
We do not detail the (hierarchic) superposition calculus that our model construction
is based on, since our assumptions are not particularly restrictive, i.e.\
we only require saturation up to redundancy w.r.t.\ (hierarchic) ordered resolution.
For a more concrete perspective, consider SUP(LA) 
\end{comment}

For pure $\HBS(\LA)$ most rules of the (hierarchic) superposition calculus
become obsolete or can be simplified.
In fact, in the $\HBS(\LA)$ case (hierarchic) superposition boils
down to (hierarchic) ordered resolution.
For a full definition of (hierarchic) superposition calculus
in the context of linear arithmetic, consider SUP(LA)\cite{DBLP:conf/frocos/AlthausKW09}.
Here, we will only define its simplified version in the form of the hierarchic resolution rule.

\begin{definition}[Hierarchic $\prec$-Resolution]
Let $\prec$ be an order on literals and
$\cc{\Lambda_1}{L_1 \lor C_1}$, $\cc{\Lambda_2}{L_2 \lor C_2}$ be constrained clauses.
The inference rule of hierarchic $\prec$-resolution is:
\[\begin{prooftree}
\hypo{\cc{\Lambda_1}{L_1 \lor C_1}}
\hypo{\cc{\Lambda_2}{L_2 \lor C_2}}
\hypo{\sigma = \mMGU(L_1, \comp(L_2))}
\infer3{(\cc{\Lambda_1, \Lambda_2}{C_1 \lor C_2}) \sigma}
\end{prooftree}\]
where $L_1$ is $\prec$-maximal in $C_1$ and $L_2$ is $\prec$-maximal in $C_2$.
\end{definition}

Note that in the resolution rule we do not enforce explicitly that the positive literal is strictly maximal.
This is possible because in the Horn case any positive literal is strictly maximal if it is maximal in the clause.

For saturation, we need a termination condition
that defines when the calculus under consideration cannot make
any further progress.
In the case of superposition, this notion is that any new inferences are \emph{redundant}.

\begin{definition}[Clause Redundancy]
A ground clause $\cc{\Lambda}{C} \in N$ is \emph{redundant} with respect to a set $N$
of ground clauses and order $\prec$ if
$N^{\prec \cc{\Lambda}{C}} \models \cc{\Lambda}{C}$.
A potentially non-ground clause $\cc{\Lambda}{C} \in N$ is \emph{redundant} with respect to a 
potentially non-ground clause set $N$ and order $\prec$ 
if for all $\cc{\Lambda'}{C'} \in \mGnd(\cc{\Lambda}{C})$ the clause $\cc{\Lambda'}{C'}$ is
redundant with respect to $\mGnd(N)$.
\end{definition}

If a clause $\cc{\Lambda}{C} \in N$ is redundant with respect to a clause set $N$, 
then it can be removed from $N$ without changing its semantics.
If $\cc{\Lambda}{C}$ is newly inferred,
then we also call it redundant if $\cc{\Lambda}{C}$ is already part of $N$.
The same cannot be said for clauses in $N$ or all clauses in $N$ would be redundant.
Determining clause redundancy is an undecidable problem~\cite{DBLP:conf/vmcai/BrombergerFW21,DBLP:conf/birthday/Weidenbach15}.
However, there are special cases of redundant clauses that can be easily checked, e.g., 
tautologies and subsumed clauses.
Redundancy also means that $\interp \models N^{\prec \cc{\Lambda}{C}}$ implies $\interp \models \cc{\Lambda}{C}$ if $\cc{\Lambda}{C}$ is redundant w.r.t.\ $N$.
We will exploit this fact in the model construction.

\begin{comment}
\begin{definition}[Tautology]
A constrained clause $\cc{\Lambda}{C}$ is a \emph{tautology} if it is satisfied independently of
the predicate interpretation (determining the truth value of $C$)
and the assignment assignment (determining the truth value of $\Lambda$).
It is therefore redundant with respect to all orders and clause sets; even the empty set.
\end{definition}
\end{comment}

\begin{definition}[Saturation]
A set of clauses $N$ is \emph{saturated up to redundancy}
with respect to some set of inference rules,
if application of any rules to
clauses in $N$ yields a clause that is redundant with respect to $N$
or is contained in $N$.
\end{definition}

\subsection{Interpretations}%

In our context, models are interpretations that satisfy
(sets of) clauses.
The standard notion of an interpretation is fairly opaque and 
interprets a predicate $P$ as the potentially 
infinite set of ground arguments that satisfy $P$.

\begin{definition}[Interpretation]
Let $P$ be a predicate symbol with $\arity(P) = n$.
Then, $P^{\interp}$
denotes the subset of $\mathcal{U}^n$ for which the \emph{interpretation} $\interp$
maps the predicate symbol $P$ to \emph{true}.
\end{definition}

Since our model construction approach manipulates
interpretations directly,
we need a notion of interpretations that always 
has a finite representation and for
which it is possible to decide (in finite time) 
whether a clause is satisfied by the interpretation.
Therefore, we rely on the notion of symbolic interpretations:

\begin{definition}[Symbolic Interpretation]
Let $x_1, x_2, \ldots$ be an infinite sequence of distinct variables, i.e.\
$x_i \ne x_j$ for all $1 \leq i < j$. 
(We assume the same sequence for all symbolic interpretations in order to prevent conflicts 
when we later combine multiple symbolic interpretations into one.)
A \emph{symbolic interpretation} $\sinterp$ is a function that maps
every predicate symbol $P$ with $\arity(P) = n$ to
a formula denoted $P^\sinterp(\vec{x})$ of finite size,
constructed using the usual boolean connectives over $\LA$ atoms, 
where the only free variables appear in $\vec{x} = (x_1, \dots, x_n)$.
The interpretation $\interp_{\sinterp}$ corresponding
to $\sinterp$ is defined by 
$P^{\interp_{\sinterp}} = \{ (\vec{x})\beta \mid \beta \models P^\sinterp(\vec{x}) \}$
and maps the predicate symbol $P$ to \emph{true}
for the subset of $\mathcal{U}^n$ which corresponds
to the solutions of $P^\sinterp(\vec{x})$.
\end{definition}

\begin{example}
Let $N$ be a clause set consisting of the clauses $0 \leq x \leq 2, 0 \leq y \leq 2 \| P(x,y)$ and $x_Q \geq x_P + 1, y_Q \geq y_P + 1 \| \neg P(x_P,y_P) \lor Q(x_Q,y_Q)$. 
An example of a symbolic interpretation $\sinterp$ that satisfies $N$, 
would be the function that maps $P$ to $P^\sinterp(x_1,x_2) = 0 \leq x_1 \leq 2 \land 0 \leq x_2 \leq 2$ and 
$Q^\sinterp(x_1,x_2) = 1 \leq x_1 \land 1 \leq x_2$.
It corresponds to the interpretation $\interp_{\sinterp}$ where $P^{\interp_{\sinterp}} = \{(a_1,a_2) \in \mathcal{U} \mid 0 \leq a_1 \leq 2 \land 0 \leq a_2 \leq 2 \}$ and 
$Q^{\interp_{\sinterp}} = \{(a_1,a_2) \in \mathcal{U} \mid 1 \leq a_1 \land 1 \leq a_2\}$.
\end{example}

The notion of symbolic interpretations is closely related to 
\emph{$\mathcal{A}$-definable models} \cite[Definition~7]{DBLP:conf/birthday/BjornerGMR15} and
\emph{constrained atomic representations} \cite[Definition~5.1,~pp.~236-237]{modelbuilding}.
Each symbolic interpretation $\sinterp(\vec{x})$ is equivalent to a constrained atomic representation 
that consists of one constraint atom $[[P(\vec{x}) : P^\sinterp(\vec{x})]]$ (written in the notation from \cite{modelbuilding})
for every predicate $P$.
Note that in this context the constraint is not just a quantifier-free conjunction of linear arithmetic atoms, 
but a linear arithmetic formula potentially containing quantifiers (although those can be eliminated with quantifier elimination techniques).

Due to the fact that each symbolic interpretation consists of a finite set of formulas of finite size, 
symbolic interpretations can be considered as finite representations.
In contrast, the standard representation of an interpretation as a potentially infinite set of ground atoms is not 
a finite representation. 
However, this also means that there are some interpretations for which no corresponding symbolic interpretation exists, 
for instance the set of prime numbers is a satisfying interpretation for $\cc{y \laeq 2}{P(y)}$, 
but not expressible as a symbolic interpretation (in $\LA$).
As we will later see, at least any saturated set of $\HBS(\LA)$ 
clauses either is unsatisfiable or has a symbolic interpretation that satisfies it (\cref{thm:horn:main}).

\begin{comment}
\begin{definition}
Let $\interp$ be an interpretation such that $P^{\interp}$ is finite for every predicate $P$.
Then, the associated symbolic interpretation $\sinterp_{\interp}$ is defined as
$P^{\sinterp_{\interp}} \defsym \bigvee_{\vec{a} \in P^\interp} \bigwedge_{1 \leq i \leq n} x_i = a_i$
for all predicate symbols $P$ of arity $n$.
\end{definition}
\end{comment}

%
The \emph{top interpretation}, denoted $\itop$,
is defined as $P^{\itop} \defsym \mathcal{U}^n$ for all predicate symbols $P$ with $\arity(P) = n$
and corresponds to
the \emph{top symbolic interpretation}, denoted $\istop$,
defined as $P^{\istop} \defsym \top$ for all predicate symbols $P$.
The \emph{bottom interpretation} (or \emph{empty interpretation}), denoted $\ibot$,
and the \emph{bottom symbolic interpretation} (or \emph{empty symbolic interpretation}), denoted $\isbot$,
are defined analogously.
The interpretation of $P$ under $\interp \cup \jnterp$ is defined as
$P^{\interp \cup \jnterp} \defsym P^\interp \cup P^\jnterp$ for every predicate $P$.
In the symbolic case,
$\sinterp \cup \sjnterp$ is defined as
$P^{\sinterp \cup \sjnterp}(\vec{x}) \defsym P^\sinterp(\vec{x}) \lor P^\sjnterp(\vec{x})$ for every predicate $P$.
%
%
\begin{comment}
\begin{definition}[Intersection of Interpretations]
Let $\interp$ and $\jnterp$ be interpretations and $P$ be a predicate symbol.
The interpretation of $P$ under $\interp \cap \jnterp$ is defined as
$P^{\interp \cap \jnterp} \defsym P^\interp \cap P^\jnterp$.
\end{definition}

\begin{definition}[Intersection of Symbolic Interpretations]
Let $\sinterp$ and $\sjnterp$ be symbolic interpretations and $P$ be a predicate symbol.
The interpretation of $P$ under $\sinterp \cap \sjnterp$ is defined as
$P^{\sinterp \cap \sjnterp} \defsym P^\sinterp(\vec{x}) \land P^\sjnterp(\vec{x})$.
\end{definition}
\end{comment}
%
%
We write
$\interp \subseteq \jnterp$ or $\interp$ is \emph{included in} $\jnterp$
(resp.\ $\interp \subset \jnterp$ or $\interp$ is \emph{strictly} included in $\jnterp$)
if $P^{\interp} \subseteq P^{\jnterp}$
(resp.\ $P^{\interp} \subset P^{\jnterp}$)
for all predicate symbols $P$.
%

\begin{comment}
\begin{proposition}
Interpretations together with their union form a commutative monoid
with the identity element $\ibot$.
\end{proposition}

\begin{proposition}
Interpretations ordered by inclusion form a complete lattice,
where the supremum is given by union
and the infimum is given by intersection.
\end{proposition}

\begin{proposition}
Interpretations ordered by inclusion form a bounded lattice,
where the maximum is given by $\itop$
and the minimum is given by $\ibot$.
\end{proposition}
\end{comment}

%

\begin{definition}[Entailment of Literal]
Let $\interp$ be an interpretation.
Given a ground literal $P(\seq{a})$,
where $a_i \in \mathcal{U}$,
we write $\interp \models P(\seq{a})$
if $(\seq{a}) \in P^{\interp}$.
Conversely, we write $\interp \nmodels P(\seq{a})$
if $(\seq{a}) \not \in P^{\interp}$.
For a non-ground literal $L$, 
we write $\interp \models L$
if for all grounding substitutions $\sigma$ for $L$,
we have $\interp \models L\sigma$.
Conversely, we write $\interp \nmodels L$,
if there exists a grounding substitution $\sigma$ for $L$,
such that $\interp \nmodels L\sigma$.
\end{definition}

We overload $\models$ for symbolic interpretations,
i.e.\ we write $\sinterp \models L$ and mean $\interp_{\sinterp} \models L$.
The following function encodes a clause as an $\LA$ formula for evaluation
under a given symbolic interpretation.

\begin{definition}[Clause Evaluation Function]\label{def:ceval}%
Let $\cc{\Lambda}{C}$ be a constrained clause 
where $C = \seqi[m]{L_\i}[1][\lor]$,
$L_i = [\lnot] P_i(\seqi[n_i]{y_{i,\i}})$
and let $\sinterp$ be a symbolic interpretation.
Then the clause evaluation function $(\cc{\Lambda}{C}\big)^{\sinterp}$ is defined as follows 
based on the definitions for $\sigma_i$ and $\phi_i$ (for $1 \leq i \leq m$):
\begin{align*}
\sigma_i \defsym \{ x_j \mapsto y_{i,j} \mid 1 \leq j \leq n_i \} \qquad
\phi_i \defsym \begin{cases}
\phantom\lnot P_i^{\sinterp} & L_i\ \text{is positive} \\
        \lnot P_i^{\sinterp} & L_i\ \text{is negative (otherwise)}
\end{cases}
\end{align*}
\begin{align*}
\big(\cc{\Lambda}{C}\big)^{\sinterp} \defsym \big(\bigwedge_{\lambda \in \Lambda} \lambda \big) \to \big(\bigvee_{i = 1}^{m} \phi_i\sigma_i \big)
\end{align*}

\end{definition}

Note that the free variables of $(\cc{\Lambda}{C})^\sinterp$ are exactly the free variables of $(\cc{\Lambda}{C})$.
Moreover, the substitutions $\sigma_i$ are necessary in the above definition 
in order to map the variables in the symbolic interpretation for the predicates $P_i^\sinterp$ to the variables that appear as arguments in the literals $P_i(\seqi[n_i]{y_{1,\i}})$.

\begin{restatable}{proposition}{thmhorneval}\label{thm:horn:eval}%
Given a constrained clause $\cc{\Lambda}{C}$ with grounding $\beta$, we have
$$\models \big(\cc{\Lambda}{C}\big)^\sinterp\beta \qquad \text{if and only if} \qquad \sinterp \models \big(\cc{\Lambda}{C}\big)\beta$$
\end{restatable}

As a corollary of the previous proposition, the entailment $\sinterp \models \cc{\Lambda}{C}$ holds if and only if the universal closure of the formula $(\cc{\Lambda}{C})^\sinterp$ is valid. 
This means that for a symbolic interpretation $\sinterp$ it is always computable whether a clause is entailed by $\sinterp$ because there are decision procedures for quantified $\LRA$, $\LQA$, and $\LIA$ formulas of finite size.

We require two functions that manipulate $\LA$-formulas directly
to express our model construction (cf.~\cref{def:delta}), i.e.\
to map solutions for a clause defined by a formula $\vars(\phi)$ 
to one atom inside the clause.
This requires from us to project away all variables in $\phi$
that appear in the clause but not in the atom.

\begin{definition}[Projection]\label{def:projection}
Let $V$ be a set of variables and $\phi$ be an $\LA$-formula.
The projection function $\pi$ is defined as follows:
$$\pi(V, \phi) \defsym \seqi{\exists x_{\i}}[1][] . \, \phi \quad \text{
where} \  \{ \seq{x} \} = \vars(\phi) \setminus V$$
\end{definition}

$\pi(V, \phi)$ is a standard projection function that binds a subset $V$ 
of the variables in the formula $\phi$ with existential quantifiers. 
Note that we also know that $\pi(V, \phi)$ is equivalent to a 
quantifier-free $\LA$ formula just over the variables 
$\seq{x}$ 
because there exist quantifier elimination algorithms for 
$\LRA$, $\LQA$, and $\LIA$~\cite{DBLP:journals/cj/LoosW93,Cooper:72a}.

A further function $\curlyvee$ is needed when we encounter literals of the form $P(x, x, \dots)$, 
i.e., where one variable is shared among two arguments.
In this case, we use $\curlyvee$ to express in our symbolic interpretation
that the equivalent argument positions must also be equivalent in our interpretation.

\begin{definition}[Sharing]\label{def:sharing}
Let $(\seq{y})$ and $(\seq{x})$ be tuples of variables with the same length.
The sharing function $\curlyvee$, which encodes variable sharing across different argument positions, is defined as follows:
$$\curlyvee \big((\seq{y}), (\seq{x})\big) \defsym \bigwedge_{1\,\leq\,i\,<\,j\,\leq\,n,\ y_i\,=\,y_j} x_i \laeq x_j$$
\end{definition}

\subsection{Consequence and Least Model}

The notion of a \emph{least model} is common in logic programming.
Horn logic programs admit a least model, which is the intersection
of all models of the program (see \cite[\S~6,~p.~36]{DBLP:books/sp/Lloyd87}).
In our context, the least model of a set of clauses $N$
is the intersection of all models of $N$.
An alternative characterization of the least model of $N$
is through the least fixed point of the one-step consequence operator,
which we define as $T_N$ for the context of $\LA$ constraints 
analogously to \cite[Section~4]{DBLP:journals/jlp/JaffarM94}.
The one-step consequence operator $T_N$ takes a set of clauses $N$ and
an interpretation $\interp$ as input
and returns an interpretation:
\begin{align*}
P^{T_N(\interp)} \defsym \left\{ (\vec{y})\beta \ \middle\vert \
\begin{array}{l}
\cc{\Lambda}{\lnot P_1(\vec{y_1}) \lor \dots \lor \lnot P_n(\vec{y_n}) \lor P(\vec{y})} \in N,\\
\models \Lambda\beta, \text{and}\ \interp \models P_i(\vec{y_i}) \beta \ \text{for}\ 1 \leq i \leq n
\end{array}
\right\}
\end{align*}
The least fixed point of this operator exists
by Tarski's Fixed Point Theorem \cite{tarski}:
Interpretations form a complete lattice under inclusion
(supremum given by union,
infimum given by intersection),
and $T_N$ is monotone.
 % end input ./prelim.tex
 %
% start input ./horn.tex
\section{Model Construction}\label{sec:horn}%

In this section we address construction of models for $\HBS(\LA)$.
Throughout this section, we consider a set of constrained Horn clauses $N$ and an order $\prec$ to be given.
Our aim is to define an interpretation
$\interpn$, such that
$$\interpn \models N \qquad \text{if $N$ is saturated and}\ \square \not \in N$$
Towards that goal, we define the operator
$\delta(\sinterp, \cc{\Lambda}{C' \lor P(\vec{y})})$.
It takes a symbolic interpretation $\sinterp$,
and a Horn clause with maximal literal $P(\vec{y})$.
It results in a symbolic interpretation that
accounts for $\cc{\Lambda}{C' \lor P(\vec{y})}$.

\begin{definition}[Production Operator]\label{def:delta}%
Let
$\cc{\Lambda}{C}$ be a constrained Horn clause, 
where $C = C' \lor P(\vec{y})$, $P(\vec{y}) \succ C'$, and
$C' = \lnot P_1(\seqi[n_1]{y_{1,\i}}) \lor \dots \lor \lnot P_m(\seqi[n_m]{y_{m,\i}})$.
Let $\sinterp$ be a symbolic interpretation,
where the free variables of $P^{\sinterp}$ are $\vec{x}$
and the free variables of $P_i^\sinterp$ are $\vec{x_i}$ (for $1 \leq i \leq m$).
Note that $n = \size{\vec{y}} = \size{\vec{x}} = \arity(P)$.
The \emph{production operator} $\delta(\sinterp, \cc{\Lambda}{C})$ 
results in a new symbolic interpretation
\begin{align*}
P^{\delta(\sinterp, \cc{\Lambda}{C})}(\vec{x}) &\ \defsym\
\Big(\pi\big(\{ \seq{y} \}, \bigwedge_{\lambda \in \Lambda} \lambda \land \bigwedge_{i = 1}^{m} (P_i^\sinterp)\sigma_i \big) \Big) \sigma \land
\curlyvee \big( \vec{y}, \vec{x} \big)\\
Q^{\delta(\sinterp, \cc{\Lambda}{C})}(\vec{z}) &\ \defsym\ \bot \qquad \text{for all}\ Q \neq P\ \text{where}\ \size{\vec{z}} = \arity(Q)
\end{align*}
where, to map variables from literal arguments to the variables appearing in the symbolic interpretation $\sinterp$ and back, 
we have the substitutions
\begin{alignat*}{8}
\sigma_{\phantom{i}} \defsym & \ \{ y' & \: \mapsto \: & x_j && \mid y' \in \{\seq{y}\} \ \text{and $j$ is the smallest index s.t. $y_j = y'$} \}\\
\sigma_i \defsym & \ \{ x_{i,j} & \: \mapsto \: & y_{i,j} && \mid 1 \leq j \leq n_i \} \qquad \text{for}\ 1 \leq i \leq m
\end{alignat*}
\end{definition}

The goal of the operator $\delta(\sinterp, \cc{\Lambda}{C})$ is to define 
an extension of the symbolic interpretation $\sinterp$ such that 
$\sinterp \cup \delta(\sinterp, \cc{\Lambda}{C})$ satisfies $\cc{\Lambda}{C}$. 
Note that $\delta$ only extends the interpretation over the strictly maximal predicate $P$.
Moreover, due to our predicate order, 
it only needs to consider the interpretation $\sinterp$ for predicates $Q$ with $Q \prec P$.
$\delta$ also satisfies the following two symmetrical properties:
On the one hand, every grounding $\tau$ of $\cc{\Lambda}{C' \lor P(\vec{y})}$ that is not yet satisfied by $\sinterp$ 
must correspond to solution $\beta$ of $P^{\delta(\sinterp, \cc{\Lambda}{C' \lor P(\vec{y})})}$ that satisfies $P(\vec{y}) \tau$.
On the other hand, every solution $\beta$ of $P^{\delta(\sinterp, \cc{\Lambda}{C' \lor P(\vec{y})})}$ must correspond to 
a grounding of $\cc{\Lambda}{C' \lor P(\vec{y})}$ that is not yet satisfied by $\sinterp$.
The first property is needed so $\sinterp \cup \delta(\sinterp, \cc{\Lambda}{C' \lor P(\vec{y})})$ satisfies $\cc{\Lambda}{C' \lor P(\vec{y})}$.
The second property is needed so we do not accidentally extend our interpretation by any solutions not needed to satisfy $\cc{\Lambda}{C' \lor P(\vec{y})}$.

Note that in the above statements $\beta$ and $\tau$ are generally not the same because 
the variables $\vec{x}$ used to define $P^{\sinterp}$ are not necessarily the same as the variables 
appearing in the clause $\cc{\Lambda}{C}$ and literal $P(\vec{y})$. 
There are three reasons for this that are handled by three different methods in our model construction:
\begin{enumerate}
\item{%
The variables in $\sinterp$ and $\cc{\Lambda}{C}$ simply do not match, e.g.\ 
in $P^{\sinterp} \defsym x_1 \laeq 0$ and $\cc{\Lambda}{C} \defsym\ \cc{y_1 > 0}{P(y_1)}$.
This is handled by the substitution $\sigma$ in $\delta$ that maps all variables in $P(\vec{y})$ 
to their appropriate variables in $P^{\sinterp}$, e.g.\ in the previous example $\sigma = \{y_1 \mapsto x_1\}$ 
and $P^{\delta(\sinterp, \cc{\Lambda}{C})} = (y_1 > 0)\sigma = x_1 > 0$.
}
\item{%
Not all variables in $\cc{\Lambda}{C}$ also appear in $P(\vec{y})$, 
e.g.\ in $P^{\sinterp} \defsym x_1 \laeq 0$ and $\cc{\Lambda}{C} \defsym\ \cc{x_1 \laeq y_1 + 1 \wedge y_1 \laeq 0}{P(x_1)}$.
This is handled in $\delta$ by the projection operator $\pi$ (\cref{def:projection}) that binds all variables that appear in $\cc{\Lambda}{C}$ but not in $P(\vec{y})$, 
e.g.\ in the previous example $P^{\delta(\sinterp, \cc{\Lambda}{C})} \defsym\ \pi(\{y_1\}, x_1 \laeq y_1 + 1 \wedge y_1 \laeq 0)$, 
where $\pi(\{y_1\}, x_1 \laeq y_1 + 1 \wedge y_1 \laeq 0) = \exists y_1. \ x_1 \laeq y_1 + 1 \wedge y_1 \laeq 0$, 
which is equivalent to $x_1 \laeq 1$.
}
\item{%
Some variables might occur in multiple argument positions, e.g.\ in $\cc{\Lambda}{C} \defsym\ \cc{\top}{P(y_1,y_1)}$.
This case is covered in $\delta$ by the sharing function $\curlyvee$ (c.f.\ \cref{def:sharing}) that 
expresses which variables in $P^{\delta(\sinterp, \cc{\Lambda}{C})}$ must map to the same value.
Continuing the example, $\curlyvee((y_1, y_1),(x_1, x_2)) = x_1 \laeq x_2$ and $P^{\delta(\sinterp, \cc{\Lambda}{C})}(x_1, x_2) = \curlyvee((y_1, y_1),(x_1, x_2))$.
}
\end{enumerate}

The parts of $P^{\delta(\sinterp, \cc{\Lambda}{C})}$ that we have not yet discussed 
are based on the fact that any constrained Horn clause $\cc{\Lambda}{C' \lor  P(\vec{y})}$ can also be written as an implication of the form
$\phi \rightarrow P(\vec{y})$, where $\phi \defsym \Lambda \land P_1(\seqi[n_1]{y_{1,\i}}) \land \dots \land P_m(\seqi[n_m]{y_{m,\i}})$ 
and
$\sinterp \nmodels \cc{\Lambda}{C'} \tau$ if and only if $\sinterp \models \phi \tau$.
This means the groundings $\tau$ of $\cc{\Lambda}{C'}$ not satisfied by $\sinterp$ are also the groundings of $\phi$ satisfied by $\sinterp$.
It is straightforward to express these groundings with a conjunctive formula based on $\Lambda$ and the $P_i^\sinterp$.
The only challenge is the reverse problem from before, i.e.\ mapping the variables of $P_i^\sinterp$ to the variables in the literals $P_i(\seqi[n_i]{y_{1,\i}})$.
This mapping is done in $\delta$ by the substitution $\sigma_i$.

Now, based on the production operator $\delta$ for one clause,
we can use an inductive definition over the order $\prec$
to define an interpretation $\sinterpn$ for all clauses in $N$.
We distinguish the following auxiliary symbolic interpretations:
$\sinterp_{\prec P}$ which captures progress up to but excluding the predicate $P$,
$\Delta_P$ which captures how $P$ should be interpreted considering $\sinterp_{\prec P}$, and
$\sinterp_{\preceq P}$ which captures progress up to and including the predicate $P$.
The symbolic interpretation $\Delta_{P}^{\cc{\Lambda}{C}}$ is the extension
of $\sinterp_{\prec P}$
w.r.t.\ the single clause $\cc{\Lambda}{C}$.

\begin{definition}[Model Construction] \label{def:modelconstruction}
Let $N$ be a finite set of constrained Horn clauses.
We define symbolic interpretations $\sinterp_{\prec P}$, $\sinterp_{\preceq P}$ and $\Delta_{P}$ for all predicates $P \in \preds(N)$ by mutual induction over $\prec$:
$$\sinterp_{\preceq P} \defsym \sinterp_{\prec P} \cup \Delta_P \qquad
\sinterp_{\prec P} \defsym \bigcup_{Q \prec P} \Delta_{Q} \qquad 
\Delta_{P} \defsym \bigcup_{\cc{\Lambda}{C' \lor P(\ast)} \in N} \Delta_{P}^{\cc{\Lambda}{C' \lor P(\ast)}}$$
$$
\Delta_{P}^{\cc{\Lambda}{C}} \defsym \begin{cases}
	\delta(\sinterp_{\prec P}, \cc{\Lambda}{C}) & \text{if} \
	P(\vec{y})\ \text{maximal in}\ C,\ \text{and}\ 
	\sinterp_{\prec P} \nmodels \cc{\Lambda}{C}
	\\
	\isbot & \text{otherwise}
\end{cases}
$$
\end{definition}

Finally, based on the above inductive definition of
$\sinterp_{\prec P}$ for every predicate symbol $P \in \preds(N)$,
we arrive at an overall interpretation for $N$.

\begin{definition}[Candidate Interpretation] \label{def:candidateinter}
The \emph{candidate interpretation}
for $N$ (w.r.t\ $\prec$), denoted $\interpn$,
is the interpretation associated with the symbolic interpretation
$\sinterpn = \bigcup_{P \in \preds(N)} \Delta_P$
where $P$ ranges over all predicate symbols occurring
in $N$.
\end{definition}

Note that $\sinterpn = \sinterp_{\preceq P}$ where $P$ is $\prec$-maximal in $\preds(N)$.
Obviously, we intend that $\sinterpn \models N$ if $N$ is saturated (\cref{thm:horn:main}).
Otherwise, i.e.\ $\sinterpn \nmodels N$, we can use our construction
to find a non-redundant inference (\cref{thm:horn:unsat}).
Consider the following two examples, demonstrating
how $\delta$ sits at the core of the aforementioned inductive definitions
of symbolic interpretations.

%
% start input ./examples.tex
\begin{example}[Dependent Interpretation]\label{ex:1}%
Assume $P \prec Q$ and consider the following set of clauses:
\begin{align*}
N &\defsym \left\{\begin{array}{lcll}
0 \leq y_1 \leq 2, 0 \leq y_2 \leq 2 &\|& \underline{P(y_1,y_2)} &\quad (C_1),\\
y_3 \geq y_1 + 1, y_4 \geq y_2 + 1 &\|& P(y_1,y_2) \to \underline{Q(y_3,y_4)} &\quad (C_2)
\end{array} \right\}
\end{align*}
Maximal literals are underlined. Since the maximal literals of $C_1$ and $C_2$ are both positive,
ordered resolution cannot be applied. The set is saturated.
Since $P$ is the $\prec$-smallest predicate we have $\sinterp_{\prec P} = \isbot$.
Applying the $\delta$ operator yields the following interpretation for $P$:
$$P^{\sinterp_{\preceq P}} = P^{\delta(\sinterp_{\prec P}, C_1)}(x_1, x_2) = 0 \leq x_1 \leq 2 \land 0 \leq x_2 \leq 2$$
Then, $Q$ is interpreted relative to $P$.
Consider the clause $C_2$:
For all solutions of its constraint
$y_3 \geq y_1 + 1, y_4 \geq y_2 + 1$
our model must also satisfy its logical part
$P(y_1,y_2) \to Q(y_3,y_4)$.
The intuition that $Q$ depends on $P$ arises
from the implication in the logical part.
Whenever the constraint of $C_2$ and $P(y_1,y_2)$ are satisfied,
$Q(y_3,y_4)$ must be satisfied.
These are exactly the points defined through $\delta(\sinterp_{\prec Q}, C_2)$,
based on
$\sinterp_{\prec Q} = \sinterp_{\preceq P} = \delta(\sinterp_{\prec P}, C_1)$:
\begin{align*}
Q^{\delta(\sinterp_{\prec Q}, C_2)}(x_1, x_2)
&= \exists z_1, z_2.\ x_1 \geq z_1 + 1 \land x_2 \geq z_2 + 1 \land 0 \leq z_1 \leq 2 \land 0 \leq z_2 \leq 2\\
&= x_1 \geq 1 \land x_2 \geq 1
\end{align*}
Whenever the conjuncts $0 \leq y_1 \leq 2$ and $0 \leq y_2 \leq 2$ are
satisfied, the premise of the implication is true,
thus there must be a solution to the interpretation of $Q$,
additionally abiding the constraint of the clause. 
Since $Q$ is $\prec$-maximal in $N$, we arrive at
$\sinterpn = \sinterp_{\preceq Q} = \sinterp_{\preceq P} \cup \delta(\sinterp_{\prec Q}, C_2) = \delta(\sinterp_{\bot}, C_1) \cup \delta(\sinterp_{\preceq P}, C_2)$.
See \cref{fig:ex:1} for a visual representation of $\sinterpn$.
\end{example}

\begin{example}[Unsaturated Clause Set]\label{ex:2}%
Assume $P \prec Q$ and consider the following set of clauses:
\begin{align*}
N &\defsym \left\{\begin{array}{llll}
\cc{y_1 < 0}{\underline{P(y_1)}} &\quad (C_1),&\qquad\quad
\cc{y_1 < 1}{\underline{Q(y_1)}} &\quad (C_3),\\
\cc{y_1 > 0}{\underline{P(y_1)}} &\quad (C_2),&\qquad\quad
\cc{y_1 \leq 0}{\underline{Q(y_1)} \to P(y_1)} &\quad (C_4)\\
\end{array} \right\}
\end{align*}
Maximal literals are underlined. Note that a resolution inference
is possible, since the maximal literals of $C_3$ and $C_4$ have opposite
polarity, use the same predicate symbol, and are trivially unifiable.
Thus, in this example we consider the effect of applying
our model construction to a clause set that is \emph{not} saturated.
Since $P$ is $\prec$-minimal, we start with the following steps:
\begin{align*}
\sinterp_{\prec P} &= \isbot &
P^{\delta(\sinterp_{\prec P}, C_1)}(x_1) &= x_1 < 0\\
& & P^{\delta(\sinterp_{\prec P}, C_2)}(x_1) &= x_1 > 0
& P^{\sinterp_{\preceq P}}(x_1) &= x_1 < 0 \lor x_1 > 0
\intertext{Next, we obtain the following results for $Q$:}
\sinterp_{\prec Q} &= \sinterp_{\preceq P} &
Q^{\delta(\sinterp_{\prec Q}, C_3)}(x_1) &= x_1 < 1 & \\
 & & Q^{\delta(\sinterp_{\prec Q}, C_4)}(x_1) &= \bot
 & Q^{\sinterp_{\preceq Q}}(x_1) &= x_1 < 1 \lor \bot = x_1 < 1
\end{align*}
See \cref{fig:ex:2} for a visual representation of $\sinterpn = \sinterp_{\preceq Q}$.
Note that $\sinterpn \nmodels C_4$, since we have
$\sinterpn \models Q(0)$ but $\sinterpn \nmodels P(0)$.
Thus, by using the constructed model, we can pinpoint
clauses that contradict that $N$ is saturated.
Applying resolution to $C_3$ and $C_4$ leads to the clause $\cc{y_1 \leq 0}{P(y_1)}$
labelled $C_5$.
If we then add $C_5$ to $N$, we instead get
$P^{\sinterp_{\preceq P}}(x_1) = x_1 < 0 \lor x_1 > 0 \lor x_1 \leq 0 = \top$.
\end{example}

\tikzset{%
cross/.style={cross out, draw=black, minimum size=2*(#1-\pgflinewidth), inner sep=0, outer sep=0, line width=0.75mm},
cross/.default={1.5mm},
grid/.pic={%
\draw[help lines, color=gray!30, dashed] (0,0) grid (3.5,3.5);

\draw[-Straight Barb] (0,0)--(3.5,0) node[right]{$x_1$};
\draw[-Straight Barb] (0,0)--(0,3.5) node[above]{$x_2$};

\foreach \x in {1,...,3} {\draw ($(\x,0) + (0,-0.05)$) -- ($(\x,0) + (0,0.05)$) node [below=0.15] {$\x$};}
\foreach \y in {1,...,3} {\draw ($(0,\y) + (-0.05,0)$) -- ($(0,\y) + (0.05,0)$) node [left =0.15] {$\y$};}
},
dots/.pic={%
\foreach \x in {1,...,3} {%
  \foreach \y in {1,...,3} {%
    \fill (\x,\y) circle[radius=0.04];
  }
}
}}

\begin{figure}[t]
\centering
\begin{subfigure}{.47\textwidth}
\centering
\begin{tikzpicture}
\pic at (0, 0) {grid} ;

\path [fill opacity=0.5, pattern={Lines[angle=+45, line width=10pt, distance=5pt]}, pattern color=black!60]
  (0,0) node [above right=0.2 and 0.2, fill opacity=1] {\Large $P$} rectangle +(2,2) ;

\path [fill opacity=0.5, pattern={Lines[angle=-45, line width=10pt, distance=5pt]}, pattern color=black!30]
  (1,1) rectangle +(2.5,2.5) ;

\draw (3,3) node [below left=0.2 and 0.2, fill opacity=1] {\Large $Q$} ;

\draw [thick, color=black] (0,0) -- (2,0) -- (2,1) -- (3,1) ;
\draw [thick, color=black] (1,3) -- (1,2) -- (0,2) -- (0,0) ;
\draw [thick, color=black, arrows = {-Stealth[scale=0.5]}] (1,3) -- (1,3.55) ;
\draw [thick, color=black, arrows = {-Stealth[scale=0.5]}] (3,1) -- (3.55,1) ;

\pic at (0, 0) {dots} ;
\end{tikzpicture}
\caption{Result of \cref{ex:1}.}
\label{fig:ex:1}
\end{subfigure}
\begin{subfigure}{.47\textwidth}
\centering
\begin{tikzpicture}
\draw[help lines, color=gray!30, dashed, ystep=0.5] (-1.5,0) grid (1.5,1.49);
\draw[Straight Barb-Straight Barb] (-1.5,0)--(1.5,0) node[right]{$x_1$} ;

\draw [thick, arrows = {Stealth[scale=0.5]-}] (-1.5,1) node [left] {$P$} -- +(1.5,0);
\draw [thick, arrows = {Stealth[scale=0.5]-}] (1.5,1) node [right] {} -- +(-1.5,0);
\draw [fill=white] (0,1) circle (0.075) ;
\draw [thick, arrows = {Stealth[scale=0.5]-}] (-1.5,0.5) node [left] {$Q$} -- +(2.5,0);
\draw [fill=white] (1,0.5) circle (0.075) ;

\foreach \x in {-1,...,1} {\draw ($(\x,0) + (0,-0.05)$) -- ($(\x,0) + (0,0.05)$) node [below=0.15] {$\x$};}

\end{tikzpicture}
\caption{Result of \cref{ex:2}.}
\label{fig:ex:2}
\end{subfigure}
\caption{Visual representation of the models resulting from \cref{ex:1,ex:2}.}
\label{fig:ex}
\end{figure}
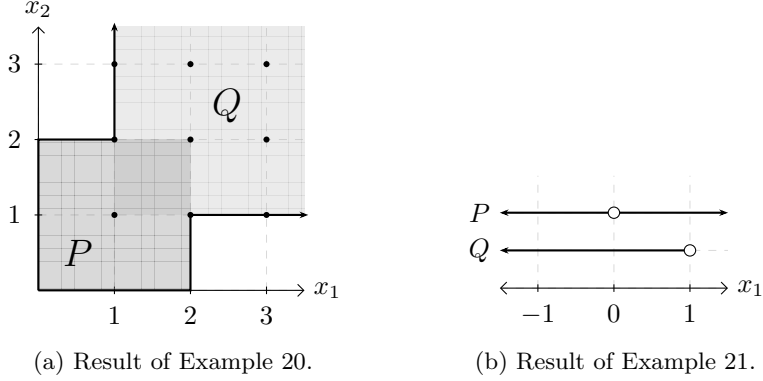
\tikzset{%
cross/.style={cross out, draw=black, minimum size=2*(#1-\pgflinewidth), inner sep=0, outer sep=0, line width=0.75mm},
cross/.default={1.5mm},
grid/.pic={%
\draw[help lines, color=gray!30, dashed] (0,0) grid (12.5,3.5);
\draw[help lines, color=gray!30, dashed] (0,-0.5) grid (12.5,-1.5);

\draw[-Straight Barb] (0,0)--(12.5,0) node[right]{$x$};
\draw[-Straight Barb] (0,0)--(0,3.5) node[above]{$y$};
\draw[-Straight Barb] (0,0)--(0,-1.5) ;

\foreach \x in {1,...,12} {\draw ($(\x,0) + (0,-0.05)$) -- ($(\x,0) + (0,0.05)$) node [below=0.15] {$\x$};}
\foreach \y in {1,...,3} {\draw ($(0,\y) + (-0.05,0)$) -- ($(0,\y) + (0.05,0)$) node [left =0.15] {$\y$};}
\foreach \y in {1,...,1} {\draw ($(0,-\y) + (-0.05,0)$) -- ($(0,-\y) + (0.05,0)$) node [left =0.15] {$-\y$};}

\draw (0,0) node [left=0.15] {$0$} ;
},
dots/.pic={%
\foreach \x in {1,...,3} {%
  \foreach \y in {1,...,3} {%
    \fill (\x,\y) circle[radius=0.04];
  }
}
}}

% end input ./examples.tex
 
In the following, we clarify some properties of the construction.
We provide an upper bound for the number of $\LA$ atoms and quantifiers in the symbolic model for $\LRA$ and $\LQA$.
Although we do not state it explicitly,
the estimate for $\LIA$ works in a similar way,
but due to the higher complexity of $\LIA$ quantifier elimination, the size of the symbolic model grows triple exponentially \cite{DBLP:journals/jcss/Oppen78}.

\begin{restatable}{proposition}{thmhorncomplexity}\label{thm:horn:complexity}
If $N$ is a finite set of $\LRA$/$\LQA$ constrained Horn clauses, and
$\sinterpn'$ the result of applying quantifier elimination to $\sinterpn$
then,
for every predicate symbol $P \in \preds(N)$,
the number of $\LA$ atoms in $P^{\sinterpn'}$ is in $O(m^{2 \cdot q^{p-1}} \cdot n^{2 \cdot q^{p-1}} \cdot (l+a^2)^{q^p})$ where
$n$ is the max.\ number of clauses with the same max.\ predicate,
$m$ is the max.\ number of non-arithmetic literals in a clause,
$l$ is the max.\ number of arithmetic literals in a clause,
$a$ is the max.\ arity of any predicate,
$p = \size{\preds(N)}$,
$q$ is the max.\ difference of variables in any clause and its positive maximal literal.
\end{restatable}
\begin{comment}
\begin{tabular}{ccl}
$n$ & $\defsym$ & $\max_{P \in \preds(N)} \size{\{\cc{\Lambda}{C} \in N \mid P\ \text{is maximal in}\ C \}}$ \\
& & (max.\ number of clauses with the same max.\ predicate) \\

$m$ & $\defsym$ & $\max_{\cc{\Lambda}{C} \in N} \size{C}$ \\
& & (max.\ number of non-arithmetic literals in a clause) \\

$l$ & $\defsym$ & $\max_{\cc{\Lambda}{C} \in N} \size{\Lambda}$ \\
& & (max.\ number of arithmetic literals in a clause) \\

$a$ & $\defsym$ & $\max_{P \in \preds(N)} \arity(P)$ \\
& & (max.\ arity of any predicate) \\

$p$ & $\defsym$ & $\size{\preds(N)}$ \\
& & (number of predicate symbols) \\

$q$ & $\defsym$ & $\max_{\cc{\Lambda}{C} \in N} \big( \vars(\cc{\Lambda}{C}) - \vars(L) \big)$ for $L \in C$ positive and maximal\\
& & (max.\ difference of variables in any clause and its pos.\ max.\ literal)
\end{tabular}
\end{comment}
\ifappendixthenelse{}{\proofthmhorncomplexity}

\begin{restatable}[Effective Construction]{corollary}{thmhorneffective}\label{thm:horn:effective}
If $N$ is a finite set of constrained Horn clauses
then for every predicate $P \in \preds(N)$,
$P^{\sinterpn}$ is a linear arithmetic formula of finite size, and
can be computed in a finite number of steps.
\end{restatable}
\ifappendixthenelse{}{\proofthmhorneffective}

We show that all points in $P^\interpn$ are necessary and justified in some sense,
that $\interpn$ is indeed a model of $N$,
and that $\interpn$ is also the least model of $N$
if $N$ is saturated.
The notion of whether a clause is productive captures
whether it contributes something to the symbolic interpretation.

\begin{definition}[Productive Clause]
Let $P$ be a predicate symbol with $\arity(P) = n$.
We say that $\cc{\Lambda}{C}$
\emph{produces} $P(\seq{a})$
if $(\seq{a}) \in P^{\Delta_{P}^{\cc{\Lambda}{C}}}$.
\end{definition}

Next, we want to formally express that every element
of the resulting interpretation is justified.
Firstly, we express that the operator $\delta$ will produce points such
that every clause is satisfied
whenever necessary, i.e.\ whenever the maximal literal of
the clause is $P(\ast)$ and the maximal literal not satisfied by $\sinterp_{\prec P}$.

\begin{restatable}{proposition}{thmhornproduces}\label{thm:horn:produces}
Let $\ccx{C}$ where
$C = C' \lor P(\vec{y})$ and
$C' \prec P(\vec{y})$.
Let $\tau$ be a grounding substitution for $\ccx{C}$.
If $\sinterp_{\prec P} \nmodels (\ccx{C})\tau$,
then
$\models \Lambda_C\tau$ and
$\sinterp_{\preceq P} \models P(\vec{y})\tau$, thus
$\sinterp_{\preceq P} \models (\ccx{C})\tau$.
\end{restatable}
\ifappendixthenelse{}{\proofthmhornproduces}

Secondly, we express that for every point in $P^\interpn$,
it is justified in the sense that there is a clause
that produced the point, i.e.\
this clause would otherwise not be satisfied by
the resulting interpretation.

\begin{restatable}{proposition}{thmhornproducer}\label{thm:horn:producer}
If $\sinterp_{\preceq P} \models P(\vec{a})$,
then there exists a clause $\ccx{C}$
where $C = C' \lor P(\vec{y})$ and
$C' \prec P(\vec{y})$,
and there exists a grounding $\tau$ for $\ccx{C}$,
such that
$P(\vec{a}) = P(\vec{y})\tau$
and
$\sinterp_{\prec P} \nmodels (\ccx{C})\tau$.
\end{restatable}
\ifappendixthenelse{}{\proofthmhornproducer}

Also, observe that once the maximal predicate $P$ of a given clause
is interpreted by $\sinterp_{\preceq P}$,
the interpretation of the clause does not change for
$\sinterp_{\preceq Q}$ where $Q \succ P$.

\begin{restatable}{corollary}{thmhorngeq}\label{thm:horn:geq}
Let $P \prec Q \preceq R$, and $P$ be maximal in clause $C$.
If $\sinterp_{\preceq P} \models \ccx{C}$ or
$\sinterp_{\prec Q} \models \ccx{C}$, then
$\sinterp_{\prec R} \models \ccx{C}$ and
$\sinterp_{\preceq R} \models \ccx{C}$.
\end{restatable}
\ifappendixthenelse{}{\proofthmhorngeq}

As a result, we know that the full model satisfies $N$, i.e., $\interpn \models N$
if every clause is satisfied at the point of the construction,
where the interpretation of its maximal predicate $P$ stays fixed.

\begin{restatable}{proposition}{thmhornmodelorproductivemodeltwo}\label{thm:horn:modelorproductivemodel2}
For every clause $\ccx{C} \in N$ with maximal predicate $P$,
if $\sinterp_{\preceq P} \models \ccx{C}$,
then
$\interpn \models N$.
\end{restatable}
\ifappendixthenelse{}{\proofthmhornmodelorproductivemodeltwo}

\ifappendixthenelse{}{Some auxiliary lemmas follow. %
% start input ./lemmas.tex
%
%
%
%
%
%
%
%

\edef\auxoffset{\thetheorem}

\setcounter{theorem}{31}

\begin{restatable}{proposition}{thmhornkeep}\label{thm:horn:keep}
Let $P$, $Q$, and $R$ be predicate symbols.
If $P \prec Q$ and $Q \preceq R$, then
$P^{\sinterp_{\preceq P}} = P^{\sinterp_{\prec Q}} = P^{\sinterp_{\preceq Q}} = P^{\sinterp_{\prec R}} = P^{\sinterp_{\preceq R}}$.
\end{restatable}
\begin{proof}
Generally, for a predicate symbol $S$, we have
$P^\Delta_S = \emptyset$ unless $S = P$
by definition of $\Delta_S$.
$\sinterp_{\prec Q}$ is defined as the union of all
$\Delta_S$ for $S \prec Q$, thus
$P^{\Delta_P} = P^{\sinterp_{\preceq P}} = P^{\sinterp_{\prec Q}}$.
$\sinterp_{\preceq Q}$ is defined as the union
of $\sinterp_{\prec Q}$ and $\Delta_Q$.
By definition of $\Delta_{Q}$, the only predicate that
$\Delta_{Q}$ may interpret as non-empty is $Q$.
In particular, i.e.\ since $P \prec Q$,
we have $P^{\Delta_{Q}} = \emptyset$
thus $P^{\sinterp_{\prec Q}} = P^{\sinterp_{\preceq Q}}$.
To see that $P^{\sinterp_{\preceq Q}} = P^{\sinterp_{\prec R}} = P^{\sinterp_{\preceq R}}$,
the same reasoning applies, i.e.\
$P^{\Delta_R} = \emptyset$ unless $P = R$.
\end{proof}

\begin{restatable}{corollary}{thmhornfixed}\label{thm:horn:fixed}
Let $P$, $Q$, and $R$ be predicate symbols.
If
\begin{enumerate}
\item $P \prec Q$, and
\item $Q \preceq R$, and
\item $\sinterp_{\preceq P} \models [\lnot]P(\vec{x})\sigma$ or
$\sinterp_{\prec Q} \models [\lnot]P(\vec{x})\sigma$,
\end{enumerate}
then
$I_{\prec R} \models [\lnot]P(\vec{x})\sigma$ and
$I_{\preceq R} \models [\lnot]P(\vec{x})\sigma$.
\end{restatable}
\begin{proof}
Corollary of \cref{thm:horn:keep}.
\end{proof}

\begin{restatable}{proposition}{thmhornfour}\label{thm:horn:4}
Let $\ccx{D}$ be a constrained Horn clause with
$D = D' \lor P(\vec{y})$.
If $\ccx{D}$ produces $P(\vec{y})\tau$,
then for all grounding substitutions $\sigma$,
such that $\sinterp_{\prec P} \nmodels (\cc{\Lambda_D}{D'})\sigma$,
for all $Q$ such that $P \preceq Q$,
we have
$\sinterp_{\preceq Q} \nmodels (\cc{\Lambda_D}{D'}) \sigma$.
\end{restatable}
\begin{proof}
Since $\ccx{D}$ produces $P(\vec{y})\tau$, all literals in $D'$
are strictly $\prec$-smaller than $P(\vec{y})$, and
$\sinterp_{\prec P} \nmodels \ccx{D}$.

Let $\sigma$ be a grounding substitution such that
$\sinterp_{\prec P} \nmodels (\cc{\Lambda_D}{D'})\sigma$.

Assume, towards a contradiction, that the proposition does not hold, i.e.\
$P \preceq Q$ and $\sinterp_{\preceq Q} \models (\cc{\Lambda_D}{D'})\sigma$.
Then we have $\models \Lambda_D \sigma$ and $\sinterp_{\preceq Q} \models ([\lnot]R(\ast))\sigma$
for some literal $[\lnot]R(\ast)$ in $D'$.

However, by assumption, all literals in $D'$ are strictly $\prec$-smaller
than $P(\vec{y})$, i.e.\ $R \prec P$.
Thus their respective interpretation is contained
in $\sinterp_{\prec P}$ by construction, i.e.\
$R^{\sinterp_{\preceq R}} = R^{\sinterp_{\prec P}} = R^{\sinterp_{\preceq Q}}$.
This contradicts $\sinterp_{\prec P} \models (\cc{\Lambda_D}{D'})\sigma$.
\end{proof}

\begin{restatable}{proposition}{thmhornsaturatedmodelorproductive}\label{thm:horn:saturatedmodelorproductive}
Let $\prec$ be a clause ordering and $N$ be a finite set of constrained Horn clauses.
If \begin{inparaenum}[(1.)]
\item{\label[premise]{thm:horn:saturatedmodelorproductive:premise:resolution}%
$N$ is saturated w.r.t.\ $\prec$-resolution, and}
\item{\label[premise]{thm:horn:saturatedmodelorproductive:premise:bot}%
there is no $\cc{\Lambda}{\bot} \in N$ where $\Lambda$ is satisfiable,}
\end{inparaenum}
then
for every clause $\ccx{C} \in N$,
$\models \ccx{C}$
or
$P$ is maximal in $C$ and
$\sinterp_{\preceq P} \models \ccx{C}$.

\end{restatable}
\begin{proof}
Assume \cref{thm:horn:saturatedmodelorproductive:premise:resolution,thm:horn:saturatedmodelorproductive:premise:bot}.
Let $\ccx{C} \in N$.
We distinguish two cases:
\begin{enumerate}
\item{$C = \bot$.
By \cref{thm:horn:saturatedmodelorproductive:premise:bot},
$\Lambda_C$ is unsatisfiable, thus $\models \ccx{C}$.
}
\item{$C = C' \lor L$ where $L$ is maximal in $C$, and $P$ is the predicate symbol associated with $L$, i.e.\ $P$ is maximal in $C$.
In case $\Lambda_C$ is unsatisfiable, $\models \ccx{C}$.
In the following, we thus consider $\Lambda_C$ satisfiable.
We distinguish two cases:
\begin{enumerate}
\item{\label[case]{horn:c1}$L = P(\vec{y})$.
We distinguish two cases:
\begin{enumerate}
\item{$\sinterp_{\prec P} \models \ccx{C}$. \cref{thm:horn:fixed} applies.
}
\item{$\sinterp_{\prec P} \nmodels \ccx{C}$. \cref{thm:horn:produces} applies.
}
\end{enumerate}
}
\item{\label[case]{thm:horn:saturatedmodelorproductive:proof:litneg}$L = \lnot P(\vec{z})$.
Assume, towards a contradiction, that the proposition does not hold, i.e.\
$\sinterp_{\preceq P} \nmodels \ccx{C}$.
Let $\sigma$ be a grounding substitution
such that
$\sinterp_{\preceq P} \nmodels (\ccx{C}) \sigma$.
It follows
that $\sinterp_{\preceq P} \nmodels (C') \sigma$ and $\sinterp_{\preceq P} \nmodels L \sigma$,
thus $\sinterp_{\preceq P} \models (P(\vec{z}))\sigma$.

Assume, without loss of generality:
\begin{itemize}
\item{\emph{minimality of $P$}:
There is no predicate symbol $Q \prec P$
such that there is a clause $\ccx{D} \in N$ with maximal predicate $Q$
and $\sinterp_{\preceq Q} \nmodels \ccx{D}$,
i.e.\
for all $Q \prec P$, $\sinterp_{\preceq Q} \models \{ \ccx{D} \in N \mid Q\ \text{maximal in}\ D \}$,
}
\item{\emph{minimality of ($\ccx{C})\sigma$}:
There is no clause $\ccx{D} \in N$ with maximal predicate $P$ and grounding $\sigma_D$, s.t.\
$(\ccx{D})\sigma_D \prec (\ccx{C})\sigma$ and $\sinterp_{\preceq P} \nmodels (\ccx{D})\sigma_D$,
i.e.\
$\sinterp_{\preceq P} \models \mGnd(\{ \ccx{D} \in N \mid P\ \text{maximal in}\ C \})^{\prec (\ccx{C})\sigma}$,
}
\item{\emph{maximality of $L \sigma$}:
there is no literal $L'$ in $C'\sigma$ s.t.\
$\lnot P(\vec{z})\sigma \prec L'$.
}
\end{itemize}

By \cref{thm:horn:producer}
there is a clause $\ccx{D} \in N$
where $D = D' \lor P(\vec{y})$, $D' \prec P(\vec{y})$,
and $\Lambda_D$ is satisfiable,
that produces $P(\vec{z})\sigma$.
Assume, without loss of generality, that the sets of variables occuring in $\ccx{C}$ and $\ccx{D}$ are disjoint,
i.e.\ $\vars(\ccx{C}) \cap \vars(\ccx{D}) = \emptyset$.
Let $\tau$ be the subsitution that maps $P(\vec{y})$ to $P(\vec{z}) \sigma$.
Since $P(\vec{y})\tau = P(\vec{z})\sigma$, there exists a most general unifier between $P(\vec{y})$ and $P(\vec{z})$, which we call $\sigma'$.
Then, since $\sigma'$ is most general,
there is a substitution $\tau'$
such that the substitution $\sigma'\tau'$ is equivalent to $\sigma$ when restricted to $\vars(\ccx{C})$ and
equivalent to $\tau$ when restricted to $\vars(\ccx{D})$.
We have $P(\vec{y})\sigma' \tau' = P(\vec{y})\tau = P(\vec{z}) \sigma$.
\newcommand{\resolventlong}{(\cc{\Lambda_C, \Lambda_D}{C' \lor D'}) \sigma'}%
\newcommand{\resolvent}{\cc{\Lambda_R}{R}}%
Consider the following $\prec$-resolution inference:
\[\begin{prooftree}
\hypo{\cc{\Lambda_C}{C' \lor \lnot P(\vec{z})}}
\hypo{\cc{\Lambda_D}{D' \lor P(\vec{y})}}
\hypo{\sigma' = \mMGU(\lnot P(\vec{z}), \comp(P(\vec{y}))}
\infer3{\resolventlong}
\end{prooftree}\]

Let $\resolventlong = \resolvent$.
Since $\models \Lambda_C \sigma$ and $\models \Lambda_D \tau$ (by \cref{thm:horn:producer}), we have $\models \Lambda_R \tau'$.
Since $\ccx{D}$ produces $P$, we have $\sinterp_{\prec P} \nmodels (\cc{\Lambda_D}{D'}) \tau$.
By \cref{thm:horn:4}, we have that,
$\sinterp_{\preceq P} \nmodels (\cc{\Lambda_D}{D'}) \tau$ and $\sinterp_{\preceq P} \nmodels (\cc{\Lambda_D}{D'}) \sigma' \tau'$.
Thus $\sinterp_{\preceq P} \nmodels \resolvent \tau'$ and $\sinterp_{\preceq P} \nmodels \resolvent$.

We distinguish two cases:

\begin{enumerate}
\item{$R = \bot$.
We distinguish two cases:
\begin{enumerate}
\item{$\cc{\Lambda_R}{\bot} \in N$. Contradicts \cref{thm:horn:saturatedmodelorproductive:premise:bot}.
}
\item{$\cc{\Lambda_R}{\bot} \not \in N$.
Then, by \cref{thm:horn:saturatedmodelorproductive:premise:resolution},
we have that $\cc{\Lambda_R}{\bot}$ is redundant, i.e.\
$N^{\prec \cc{\Lambda_R}{\bot}} \models \cc{\Lambda_R}{\bot}$.
However, since there is no clause that is $\prec$-smaller
than $\cc{\Lambda_R}{\bot}$,
we have $N^{\prec \cc{\Lambda_R}{\bot}} = \emptyset$,
which contradicts \cref{thm:horn:saturatedmodelorproductive:premise:resolution}.
}
\end{enumerate}
}
\item{
$R \ne \bot$.
$\lnot P(\vec{z})\sigma$ occurs less often in $R\tau'$
than it occurs in $C\sigma$.
The reason being that
the number of occurences in $C'\sigma'$ is one less
than the number of occurences in $C\sigma'$,
while there are no occurences in $D'\sigma'$
since we know that $D'\sigma' \prec P(\vec{y})\sigma'$ and $D\sigma' \preceq P(\vec{y})\sigma'$,
and $P(\vec{y})\sigma' \prec \lnot P(\vec{z})\sigma$.
By maximality of $\lnot P(\vec{z})\sigma$,
also $\lnot P(\vec{y})\tau' \preceq R\tau'$.
Therefore we have $\resolvent \prec \ccx{C}$.
By minimality of $(\ccx{C})\sigma$, \cref{horn:c1}, and \cref{thm:horn:geq},
we have $\sinterp_{\preceq P} \models \mGnd(N)^{\prec (\ccx{C})\sigma}$.
We distinguish two cases:
\begin{enumerate}
\item{
$\resolvent \in N$.
Contradicts minimality of $(\ccx{C})\sigma$.
Contradicts $\sinterp_{\preceq P} \models \mGnd(N)^{\prec (\ccx{C})\sigma}$ since $(\resolvent)\tau' \prec (\ccx{C})\sigma$.
}
\item{
$\resolvent \not \in N$. By assumption of \cref{thm:horn:saturatedmodelorproductive:premise:resolution}, $\resolvent$ is redundant w.r.t.\ $N$ and $\prec$, i.e.\ $N^{\prec \resolvent} \models \resolvent$.
$\sinterp_{\preceq P} \models \mGnd(N)^{\prec (\ccx{C})\sigma}$
contradicts
$\mGnd(N)^{\prec (\resolvent)\tau'} \models (\resolvent)\tau'$
and $N^{\prec \resolvent} \models \resolvent$.
Therefore $\resolvent$ is not redundant, which contradicts \cref{thm:horn:saturatedmodelorproductive:premise:resolution}.
}
\end{enumerate}
}
\end{enumerate}
}
\end{enumerate}
}
\end{enumerate}
\end{proof}

\setcounter{theorem}{\auxoffset} % end input ./lemmas.tex
}

With the above propositions
\ifappendixthenelse{(and some auxiliary properties that can be found in \cite{DBLP:journals/corr/abs-2305-05064})}{}
we show that indeed $\interp_N \models N$ if $N$ is saturated and does not contain the empty clause.

\begin{restatable}{theorem}{thmhornmain}\label{thm:horn:main}
Let $\prec$ be a clause ordering and $N$ be a set of constrained Horn clauses.
If \begin{inparaenum}[(1.)]
\item{%
$N$ is saturated w.r.t.\ $\prec$-resolution, and}
\item{%
$\square \not \in N$,}
\end{inparaenum}
then $\interpn \models N$.
\end{restatable}
\ifappendixthenelse{}{\proofthmhornmain}

For clauses with positive maximal literal,
the fact that they are satisfied by $\interpn$
follows from \cref{thm:horn:produces}.
For clauses with maximal literal $\lnot P(\ast)$,
we prove this theorem by contradiction:
If there is a minimal clause $\ccx{C}$ such
that $\sinterpn \nmodels \ccx{C}$.
We can then exploit \cref{thm:horn:producer}
to find the smallest clause $\ccx{D}$
that produced the respective instance $P(\vec{a})$.
Applying hierarchic $\prec$-resolution to $\ccx{C}$ and $\ccx{D}$ then yields 
a non-redundant clause.
This idea then leads to the following theorem.

\begin{restatable}{corollary}{thmhornunsat}\label{thm:horn:unsat}
Let $\prec$ be a clause ordering and $N$ be a set of constrained Horn clauses.
If \begin{inparaenum}[(1.)]
\item{%
$\interpn \nmodels N$, and}
\item{%
$\square \not \in N$,}
\end{inparaenum}
then there exist two clauses $\ccx{C}$, $\ccx{D} \in N$ such that:
\begin{inparaenum}[(1.)]
\item{%
$\ccx{C}$ is the smallest clause not satisfied by $\interpn$, i.e.
there exists a grounding $\tau$ such that
$\interpn \nmodels (\ccx{C})\tau$, but
there does not exist a clause $\ccx{C'} \in N$ with grounding $\tau'$, such that $\interpn \nmodels (\ccx{C'})\tau'$ and $(\ccx{C'})\tau' \prec (\ccx{C})\tau$,
}
\item{%
$\lnot P(\vec{a})$ is the maximal literal of $(\ccx{C})\tau$,
}
\item{%
$\ccx{D}$ is the minimal clause that produces $P(\vec{a})$,
}
\item{%
$\prec$-resolution is applicable to $\ccx{C}$ and $\ccx{D}$, and
}
\item{%
the resolvent of $\ccx{C}$ and $\ccx{D}$ is not redundant w.r.t.\ $N$.
}
\end{inparaenum}
\end{restatable}
\ifappendixthenelse{}{\proofthmhornunsat}

Additionally, we show that $\interp_N$ is the least model of $N$,
establishing a connection between our approach and the literature
on constrained Horn clauses (see \cite[Section~4]{DBLP:journals/jlp/JaffarM94} and
\cite[Section~2.4.1]{DBLP:journals/tplp/AngelisFGHPP22}) and logic programming
(see \cite[\S~6,~p.~37]{DBLP:books/sp/Lloyd87}).

\begin{restatable}{theorem}{thmhornlm}\label{thm:horn:lm}
$\interpn$ is the least model of $N$.
\end{restatable}
\ifappendixthenelse{}{\proofthmhornlm}

\begin{comment}
Further, we establish a connection between the $T_N$ operator (known as $T_P$ in logic programming)
and superposition.
We can show that the the fixed point of the one-step consequence operator
is reached in a finite amount steps
if and only if
superposition saturates in finite amount of steps.

\begin{restatable}{theorem}{thmhornterm}\label{thm:horn:term}
The least fixed point of $T_N$ is reached in finitely many steps
if and only if
hierarchic superposition saturates on $N$ in finitely many steps.
\end{restatable}
\end{comment}

Fermüller and Leitsch define four postulates (see \cite{DBLP:journals/logcom/FermullerL96} as cited in \cite[Section~5.1,~p.~234]{modelbuilding}) regarding \emph{automated model building}.
In the following, we instantiate the postulates for our setting.
By $\sinterps(N)$ we denote the set of all symbolic interpretations of the set of constrained Horn clauses $N$.
We argue how our approach satisfies all postulates, one by one:

\begin{description}
\item[Uniqueness]{%
\emph{Each element of $\sinterps(N)$ specifies a single interpretation of $N$.}\\
We have shown (cf.\ \cref{thm:horn:lm}) that $\interp_N$, the model represented by $\sinterp_N$, is the least model of $N$, which is unique.
}

\item[Atom Test]{%
\emph{There exists a fast procedure to evaluate arbitrary ground atoms over $\preds(N)$ in the interpretation defined by a $\sinterp$ in $\sinterps(N)$.}\\
This is a special case of clause evaluation (cf.\ \cref{thm:horn:eval}): A ground atom $P(\vec{t})$ is true in $\sinterp$ if and only if $\models P^{\sinterp}(\vec{x})\{ x_i \mapsto t_i \mid 1 \leq i \leq \size{\vec{x}} = \size{\vec{t}}\}$.
Fulfillment of this property thus hinges on the meaning of \enquote{fast}.
We consider methods for evaluating formulas of $\LA$ against points to be fast.
}

\item[Formula Evaluation]{%
\emph{There exists an algorithm deciding the truth values of arbitrary formulas
in interpretations defined by $\sinterp \in \sinterps(N)$.}\\
\cref{thm:horn:eval} states that evaluating a constrained clause $\cc{\Lambda}{C}$ is achieved by evaluating the universal closure of $(\cc{\Lambda}{C})^{\sinterp}$, which is decided by
quantifier elimination algorithms for $\LRA$, $\LQA$, and $\LIA$~\cite{DBLP:journals/cj/LoosW93,Cooper:72a}.
For sets of clauses, evaluate each clause individually and combine the results conjunctively.
}

\item[Equivalence Test]{%
\emph{There exists an algorithm which decides whether two representations $\sinterp_1$ and $\sinterp_2$ in $\sinterps(N)$ describe the same interpretation.}\\
$\sinterp_1$ and $\sinterp_2$ describe the same interpretation if and only if for each predicate $P \in \preds(N)$ of arity $n$, we have $\forall x_1 \dots \forall x_n . \, P^{\sinterp_1}(\vec{x}) \leftrightarrow P^{\sinterp_2}(\vec{x})$.
}
\end{description} % end input ./horn.tex
 %
% start input ./conclusion.tex
\section{Conclusion}\label{sec:conclusion}%

We have presented the first model construction approach to
Horn clauses with linear arithmetic constraints based on hierarchic ordered resolution, (cf. \cref{def:candidateinter}).
The linear arithmetic constraints may range over the reals, rationals, or integers.
The computed model is the canonical least model of the
saturated Horn clause set (cf.\ \cref{thm:horn:lm}).  Clauses can be effectively evaluated
with respect to the model (cf.\ \cref{thm:horn:eval}). This offers a way to explore the properties
of a saturated clause set, e.g., if the set represents a failed refutation
attempt.

\paragraph{Future Work}\label{sec:future}%
It is straightforward to see that any symbolic $\LQA$ model is also a symbolic $\LRA$ model.
(This holds due to convexity of conjunctions of ground $\LQA$ atoms.)
So even if the axiom of choice is not assumed, 
there is an alternative way to obtain a model for a $\HBS(\LRA)$ clause set:
Simply treat it as an $\HBS(\LQA)$ clause set, saturate it and construct its model based 
on $\HBS(\LQA)$.

In this work, we restrict ourselves to only one sort $\LA$ per set of clauses.
An extension to a many-sorted setup,
e.g.\ including first-order variables with sort $\FolS$
is possible.
This can even be simulated, by encoding first-order constants
as concrete natural numbers
via a bijection to $\mathbb{N}$, since $\mathbb{N} \subset \mathcal{U}$.
By not placing any arithmetic constraints on the variables used
for the encoding, it can be read off and mapped back from the resulting model.

One obvious challenge is relaxation of the restriction to Horn clauses. With respect
to ordered resolution saturation there is typically no difference in the sense that if a Horn fragment can always be
finitely saturated, so can the non-Horn fragment be.
However, our proposed ordering for the model construction at the granularity of predicate symbols will not suffice in this general case, 
and the key to overcome this challenge seems to be the appropriate treatment of clauses with
maximal literals of the same predicate.
Backtracking on the selection of literals might also be sufficient.

The approach we presented does not exploit features of linear arithmetic beyond equality and 
the existence of a well-founded order for the underlying universe $\mathcal{U}$.
The results may therefore be adapted to other constraint domains such as non-linear
arithmetic.

\paragraph{Acknowledgements} We thank our reviewers for their constructive comments.
 % end input ./conclusion.tex
 
\clearpage
\bibliographystyle{splncs04}

\begin{thebibliography}{10}
\providecommand{\url}[1]{\texttt{#1}}
\providecommand{\urlprefix}{URL }
\providecommand{\doi}[1]{https://doi.org/#1}

\bibitem{DBLP:conf/frocos/AlthausKW09}
Althaus, E., Kruglov, E., Weidenbach, C.: Superposition modulo linear
  arithmetic {SUP(LA)}. In: FroCoS 2009. LNCS, vol.~5749, pp. 84--99. Springer
  (2009). \doi{10.1007/978-3-642-04222-5_5}

\bibitem{DBLP:conf/kgc/BachmairGW93}
Bachmair, L., Ganzinger, H., Waldmann, U.: Superposition with simplification as
  a desision procedure for the monadic class with equality. In: Computational
  Logic and Proof Theory, Third Kurt G{\"{o}}del Colloquium, KGC'93, Brno,
  Czech Republic, August 24-27, 1993, Proceedings. LNCS, vol.~713, pp. 83--96.
  Springer (1993). \doi{10.1007/BFb0022557}

\bibitem{DBLP:journals/aaecc/BachmairGW94}
Bachmair, L., Ganzinger, H., Waldmann, U.: Refutational theorem proving for
  hierarchic first-order theories. AAECC  \textbf{5},  193--212 (1994).
  \doi{10.1007/BF01190829}

\bibitem{DBLP:journals/jacm/BasinG01}
Basin, D.A., Ganzinger, H.: Automated complexity analysis based on ordered
  resolution. JACM  \textbf{48}(1),  70--109 (2001).
  \doi{10.1145/363647.363681}

\bibitem{DBLP:conf/lpar/BaumgartnerFT08}
Baumgartner, P., Fuchs, A., Tinelli, C.: {(LIA)} - model evolution with linear
  integer arithmetic constraints. In: LPAR 2008. LNCS, vol.~5330, pp. 258--273.
  Springer (2008). \doi{10.1007/978-3-540-89439-1_19}

\bibitem{DBLP:conf/birthday/0001W19}
Baumgartner, P., Waldmann, U.: Hierarchic superposition revisited. In:
  Description Logic, Theory Combination, and All That - Essays Dedicated to
  Franz Baader on the Occasion of His 60th Birthday. LNCS, vol. 11560, pp.
  15--56. Springer (2019). \doi{10.1007/978-3-030-22102-7_2}

\bibitem{DBLP:conf/birthday/BjornerGMR15}
Bj{\o}rner, N.S., Gurfinkel, A., McMillan, K.L., Rybalchenko, A.: Horn clause
  solvers for program verification. In: Fields of Logic and Computation {II} -
  Essays Dedicated to Yuri Gurevich on the Occasion of His 75th Birthday. LNCS,
  vol.~9300, pp. 24--51. Springer (2015). \doi{10.1007/978-3-319-23534-9_2}

\bibitem{DBLP:conf/tacas/BrombergerDFFGK22}
Bromberger, M., Dragoste, I., Faqeh, R., Fetzer, C., Gonz{\'{a}}lez, L.,
  Kr{\"{o}}tzsch, M., Marx, M., Murali, H.K., Weidenbach, C.: A sorted datalog
  hammer for supervisor verification conditions modulo simple linear
  arithmetic. In: {TACAS 2022} as part of {ETAPS 2022}. LNCS, vol. 13243, pp.
  480--501. Springer (2022). \doi{10.1007/978-3-030-99524-9_27}

\bibitem{DBLP:conf/frocos/BrombergerDFFKW21}
Bromberger, M., Dragoste, I., Faqeh, R., Fetzer, C., Kr{\"{o}}tzsch, M.,
  Weidenbach, C.: A datalog hammer for supervisor verification conditions
  modulo simple linear arithmetic. In: FroCoS 2021. LNCS, vol. 12941, pp.
  3--24. Springer (2021). \doi{10.1007/978-3-030-86205-3_1}

\bibitem{DBLP:conf/vmcai/BrombergerFW21}
Bromberger, M., Fiori, A., Weidenbach, C.: Deciding the bernays-schoenfinkel
  fragment over bounded difference constraints by simple clause learning over
  theories. In: VMCAI 2021. LNCS, vol. 12597, pp. 511--533. Springer (2021).
  \doi{10.1007/978-3-030-67067-2_23}

\bibitem{DBLP:conf/cade/BrombergerLW22}
Bromberger, M., Leutgeb, L., Weidenbach, C.: An efficient subsumption test
  pipeline for {BS(LRA)} clauses. In: IJCAR 2022. LNCS, vol. 13385, pp.
  147--168. Springer (2022). \doi{10.1007/978-3-031-10769-6_10}

\bibitem{DBLP:journals/corr/abs-2305-05064}
Bromberger, M., Leutgeb, L., Weidenbach, C.: Symbolic model construction for
  saturated constrained horn clauses. arXiv  (2023).
  \doi{10.48550/arXiv.2305.05064}

\bibitem{modelbuilding}
Caferra, R., Leitsch, A., Peltier, N.: Automated Model Building, APLS, vol.~31.
  Springer (2004). \doi{10.1007/978-1-4020-2653-9}

\bibitem{Cooper:72a}
Cooper, D.C.: Theorem proving in arithmetic without multiplication. Machine
  Intelligence  \textbf{7},  91--99 (1972)

\bibitem{DBLP:journals/tplp/AngelisFGHPP22}
{De Angelis}, E., Fioravanti, F., Gallagher, J.P., Hermenegildo, M.V.,
  Pettorossi, A., Proietti, M.: Analysis and transformation of constrained horn
  clauses for program verification. TPLP  \textbf{22}(6),  974--1042 (2022).
  \doi{10.1017/S1471068421000211}

\bibitem{Downey1972}
Downey, P.J.: Undecidability of presburger arithmetic with a single monadic
  predicate letter. Tech. rep., Center for Research in Computer Technology,
  Harvard University (1972)

\bibitem{DBLP:conf/cav/FedyukovichZG18}
Fedyukovich, G., Zhang, Y., Gupta, A.: Syntax-guided termination analysis. In:
  CAV 2018. LNCS, vol. 10981, pp. 124--143. Springer (2018).
  \doi{10.1007/978-3-319-96145-3_7}

\bibitem{Feferman1964}
Feferman, S.: Some applications of the notions of forcing and generic sets.
  Fundamenta Mathematicae  \textbf{56}(3),  325--345 (1964),
  \url{http://eudml.org/doc/213821}

\bibitem{DBLP:journals/logcom/FermullerL96}
Ferm{\"{u}}ller, C.G., Leitsch, A.: Hyperresolution and automated model
  building. LOGCOM  \textbf{6}(2),  173--203 (1996).
  \doi{10.1093/logcom/6.2.173}

\bibitem{DBLP:journals/igpl/FermullerL98}
Ferm{\"{u}}ller, C.G., Leitsch, A.: Decision procedures and model building in
  equational clause logic. IGPL  \textbf{6}(1),  17--41 (1998).
  \doi{10.1093/jigpal/6.1.17}

\bibitem{DBLP:journals/corr/abs-2003-04627}
Fiori, A., Weidenbach, C.: {SCL} with theory constraints. arXiv  (2020),
  \url{https://arxiv.org/abs/2003.04627}

\bibitem{DBLP:journals/tplp/GangeNSSS15}
Gange, G., Navas, J.A., Schachte, P., S{\o}ndergaard, H., Stuckey, P.J.: Horn
  clauses as an intermediate representation for program analysis and
  transformation. TPLP  \textbf{15}(4-5),  526--542 (2015).
  \doi{10.1017/S1471068415000204}

\bibitem{DBLP:conf/lics/GanzingerN99}
Ganzinger, H., de~Nivelle, H.: A superposition decision procedure for the
  guarded fragment with equality. In: 14th LICS, 1999. pp. 295--303. {IEEE}
  Computer Society (1999). \doi{10.1109/LICS.1999.782624}

\bibitem{DBLP:conf/pldi/GrebenshchikovLPR12}
Grebenshchikov, S., Lopes, N.P., Popeea, C., Rybalchenko, A.: Synthesizing
  software verifiers from proof rules. In: PLDI. pp. 405--416. {ACM} (2012).
  \doi{10.1145/2254064.2254112}

\bibitem{DBLP:conf/sat/HoderB12}
Hoder, K., Bj{\o}rner, N.S.: Generalized property directed reachability. In:
  SAT 2012. LNCS, vol.~7317, pp. 157--171. Springer (2012).
  \doi{10.1007/978-3-642-31612-8_13}

\bibitem{DBLP:journals/corr/HorbachVW17}
Horbach, M., Voigt, M., Weidenbach, C.: The universal fragment of presburger
  arithmetic with unary uninterpreted predicates is undecidable. arXiv  (2017),
  \url{http://arxiv.org/abs/1703.01212}

\bibitem{DBLP:journals/jlp/JaffarM94}
Jaffar, J., Maher, M.J.: Constraint logic programming: {A} survey. JLP
  \textbf{19/20},  503--581 (1994). \doi{10.1016/0743-1066(94)90033-7}

\bibitem{DBLP:conf/cav/KomuravelliGC14}
Komuravelli, A., Gurfinkel, A., Chaki, S.: Smt-based model checking for
  recursive programs. In: {CAV 2014} as part of {VSL 2014}. LNCS, vol.~8559,
  pp. 17--34. Springer (2014). \doi{10.1007/978-3-319-08867-9_2}

\bibitem{DBLP:conf/csl/KorovinV07}
Korovin, K., Voronkov, A.: Integrating linear arithmetic into superposition
  calculus. In: CSL 2007. LNCS, vol.~4646, pp. 223--237. Springer (2007).
  \doi{10.1007/978-3-540-74915-8_19}

\bibitem{DBLP:phd/dnb/Kruglov13}
Kruglov, E.: Superposition modulo theory. Ph.D. thesis, Saarland University
  (2013), \url{http://scidok.sulb.uni-saarland.de/volltexte/2013/5559/}

\bibitem{DBLP:books/sp/Lloyd87}
Lloyd, J.W.: Foundations of Logic Programming, 2nd Edition. Springer (1987).
  \doi{10.1007/978-3-642-83189-8}

\bibitem{DBLP:journals/cj/LoosW93}
Loos, R., Weispfenning, V.: Applying linear quantifier elimination. The
  Computer Journal  \textbf{36}(5),  450--462 (1993).
  \doi{10.1093/comjnl/36.5.450}

\bibitem{DBLP:journals/tplp/Lopez-GarciaDKL18}
L{\'{o}}pez{-}Garc{\'{\i}}a, P., Darmawan, L., Klemen, M., Liqat, U., Bueno,
  F., Hermenegildo, M.V.: Interval-based resource usage verification by
  translation into horn clauses and an application to energy consumption. TPLP
  \textbf{18}(2),  167--223 (2018). \doi{10.1017/S1471068418000042}

\bibitem{DBLP:conf/cav/McMillan14}
McMillan, K.L.: Lazy annotation revisited. In: {CAV 2014} as part of {VSL
  2014}. LNCS, vol.~8559, pp. 243--259. Springer (2014).
  \doi{10.1007/978-3-319-08867-9_16}

\bibitem{DBLP:journals/tplp/MesnardPV20}
Mesnard, F., Payet, {\'{E}}., Vidal, G.: Concolic testing in {CLP}. TPLP
  \textbf{20}(5),  671--686 (2020). \doi{10.1017/S1471068420000216}

\bibitem{DBLP:journals/jcss/Oppen78}
Oppen, D.C.: A 2{\^{}}2{\^{}}2{\^{}}pn upper bound on the complexity of
  presburger arithmetic. JCSS  \textbf{16}(3),  323--332 (1978).
  \doi{10.1016/0022-0000(78)90021-1}

\bibitem{DBLP:conf/lpar/Rummer08}
R{\"{u}}mmer, P.: A constraint sequent calculus for first-order logic with
  linear integer arithmetic. In: LPAR 2008. LNCS, vol.~5330, pp. 274--289.
  Springer (2008). \doi{10.1007/978-3-540-89439-1_20}

\bibitem{DBLP:journals/toplas/SpotoMP10}
Spoto, F., Mesnard, F., Payet, {\'{E}}.: A termination analyzer for java
  bytecode based on path-length. TOPLAS  \textbf{32}(3),  8:1--8:70 (2010).
  \doi{10.1145/1709093.1709095}

\bibitem{tarski}
Tarski, A.: A lattice-theoretical fixpoint theorem and its applications.
  Pacific Journal of Mathematics  \textbf{5}(2),  285 -- 309 (1955).
  \doi{pjm/1103044538}

\bibitem{DBLP:conf/birthday/Weidenbach15}
Weidenbach, C.: Automated reasoning building blocks. In: Correct System Design
  - Symposium in Honor of Ernst-R{\"{u}}diger Olderog on the Occasion of His
  60th Birthday, Oldenburg, Germany, September 8-9, 2015. Proceedings. LNCS,
  vol.~9360, pp. 172--188. Springer (2015). \doi{10.1007/978-3-319-23506-6_12}

\end{thebibliography}

\end{document}